\documentclass[conference]{IEEEtran}
\IEEEoverridecommandlockouts
\usepackage{cite}
\usepackage{amsmath,amssymb,amsfonts}
\usepackage{algorithmicx}
\usepackage{graphicx}
\usepackage{textcomp}
\usepackage{xcolor}
\def\BibTeX{{\rm B\kern-.05em{\sc i\kern-.025em b}\kern-.08em
    T\kern-.1667em\lower.7ex\hbox{E}\kern-.125emX}}

\usepackage[hyphens]{url}

\usepackage{subcaption} 
\usepackage{tabularx}
\usepackage{multirow}
\usepackage{amsmath}
\usepackage{pifont}
\usepackage{algorithm}
\usepackage{algpseudocode}
\usepackage{amsfonts}
\usepackage{soul}
\usepackage{booktabs}
\usepackage{pifont}
\definecolor{commentcolor}{RGB}{0,128,0}
\definecolor{commentcolor_red}{RGB}{210,0,0}

\begin{document}

\title{BitDecoding: Unlocking Tensor Cores for Long-Context LLMs with Low-Bit KV Cache
}

\author{
Dayou Du$^{1\dagger}$\thanks{$^{\dagger}$Work partially done during an internship at Microsoft Research.}%
,\; Shijie Cao$^{2*}$\thanks{$^{*}$Corresponding author.}%
,\; Jianyi Cheng$^{1}$,\; Luo Mai$^{1}$,\; Ting Cao$^{3}$,\; Mao Yang$^{2}$\\
$^{1}$University of Edinburgh
$^{2}$Microsoft Research\\
$^{3}$Institute for AI Industry Research (AIR), Tsinghua University\\
{\{dayou.du, jianyi.cheng, luo.mai\}@ed.ac.uk,\; \{shijiecao, maoyang\}@microsoft.com,\; tingcao@mail.tsinghua.edu.cn}
}

\maketitle

\begin{abstract}
The growth of long-context Large Language Models (LLMs) significantly increases memory and bandwidth pressure during autoregressive decoding due to the expanding Key–Value (KV) cache. While accuracy-preserving KV-cache quantization (e.g., 4-bit or 2-bit) reduces memory footprint, existing systems decode inefficiently by relying solely on CUDA cores, underutilizing Tensor Cores—the dominant compute resource on GPUs.

We present BitDecoding, the first inference system to efficiently decode low-bit KV caches by cooperatively leveraging CUDA cores and Tensor Cores. BitDecoding smartly induces Tensor-Core-friendly layouts, introduces warp-level dequantization parallelism, and provides unified system support through query transformation, high-performance tensor- and channel-wise quantization, and a software-pipelined dequantization kernel enabling mixed-precision execution. Architecture-aware optimizations further leverage Hopper’s warpgroup tensor instructions and Blackwell’s NVFP4 (MXFP4) tensor formats.

Evaluated on Blackwell, Hopper, and Ampere GPUs, BitDecoding achieves an average 7.5$\times$ decoding speedup over FP16 FlashDecoding-v2, up to 8.6$\times$ on Blackwell with NVFP4, and up to 4.3$\times$ over state-of-the-art approaches. On LLaMA-3.1-8B with a 128K context, BitDecoding reduces single-batch decoding latency by 3$\times$. BitDecoding is open-sourced at \url{https://github.com/OpenBitSys/BitDecoding}.
\end{abstract}


\section{Introduction}


The ability of Large Language Models (LLMs) to process \textbf{long contexts}~\cite{yarn,longrope,gemini} has unlocked new capabilities, such as book summarization \cite{booookscore}, multi-modal understanding \cite{mm}, and test-time scaling \cite{deepseek-r1, openai_o3_mini}. However, these advancements come with significant memory and computational challenges, primarily due to the large size of the Key-Value (KV) cache in long-context scenarios. During autoregressive decoding, LLMs must repeatedly access this growing cache for each generated token, which increases memory usage and slows down decoding. The problem worsens with larger batch sizes, as the KV cache scales linearly with the number of concurrent queries. For example, a 7B model requires approximately 14GB GPU memory for its parameters, but with a 32K context length and a batch size of 8, the KV cache alone consumes 128GB GPU memory \cite{kvquant}, creating a significant memory bottleneck.

To address this growing bottleneck, \textbf{KV cache quantization} has emerged as a promising solution. By reducing the bit-width of the KV cache, quantization lowers memory overhead and improves overall efficiency. Recent quantization algorithms have shown that low-bit KV cache can retain high accuracy. 
QServe~\cite{qserve} demonstrates 4-bit KV cache improves throughput on models like LLaMA-3 and Qwen-1.5 while maintaining strong accuracy, even together with 4-bit weight and 8-bit activation. 
Further research~\cite{kivi, gear, rotatekv} shows that 2-bit KV cache can achieve near fp16 accuracy.
Kivi~\cite{kivi}, for instance, incurs only a 0.6\% accuracy drop on LongBench~\cite{bai2024longbench} with a 2-bit KV cache on LLaMA-2-7B-Chat.
Recent studies~\cite{1bitkv, asymkv} explore 1-bit quantization for KV cache, maintaining acceptable accuracy under specific conditions.
These results confirm that KV cache quantization strikes an effective balance between efficiency and accuracy, making it viable for long-context LLM deployment.

\begin{figure}[t]
    \centering
    \begin{subfigure}[b]{0.49\linewidth}
        \centering
        \raisebox{2mm} {\includegraphics[width=\linewidth]{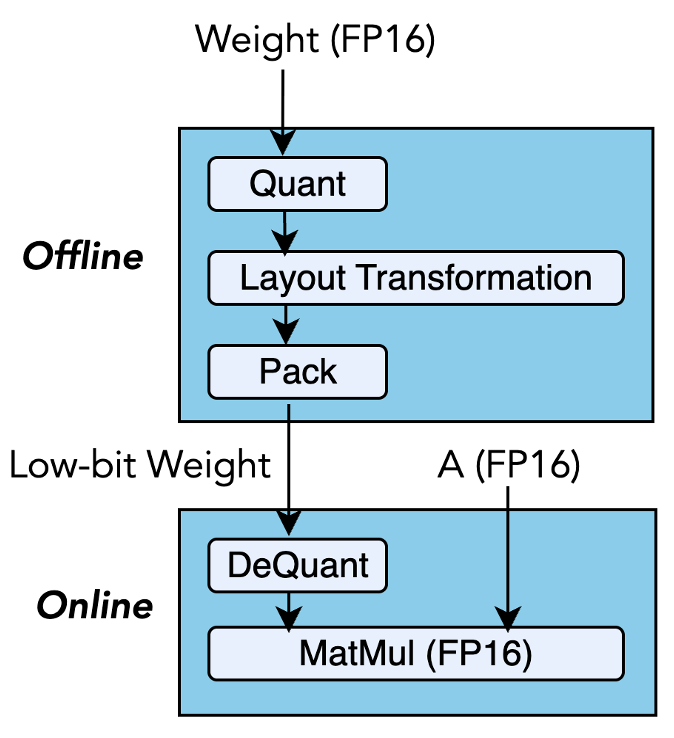}}
        \caption{Low-bit weight}
        \label{fig:Transformation}
    \end{subfigure}
    \hfill
    \begin{subfigure}[b]{0.4\linewidth}
        \centering
        {\includegraphics[width=\linewidth]{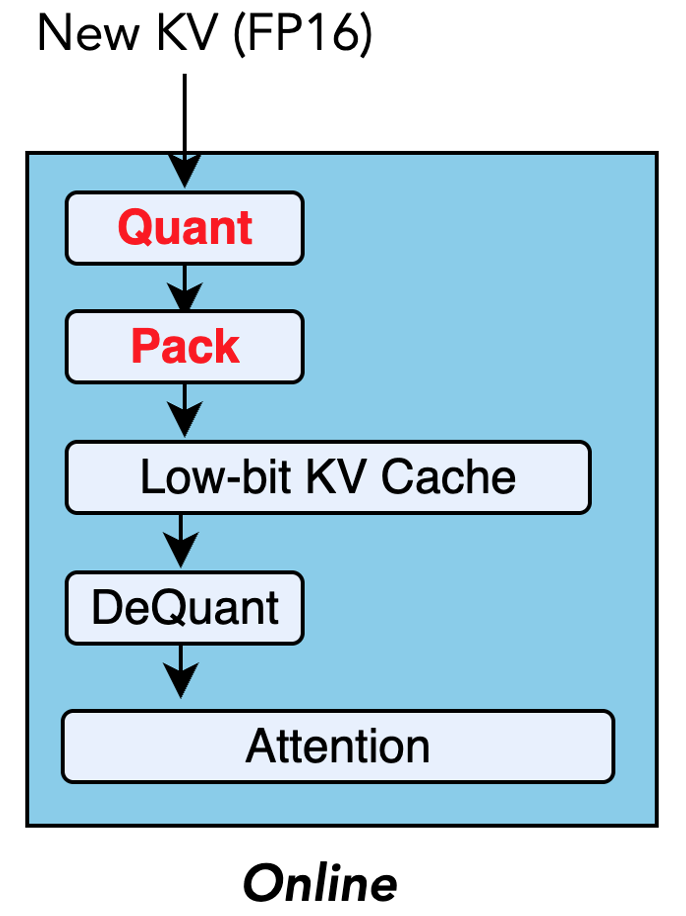}}
        \caption{Low-bit KV cache}
        \label{fig:gqa_performance}
    \end{subfigure}
    \caption{Comparison of mixed-precision matrix multiplication for low-bit weight and low-bit KV cache. (a) Quantized weights can be preprocessed offline. (b) KV cache requires online quantization and packing for each newly generated token.}
    \label{fig:diff}
\end{figure}

\textit{Despite the memory savings, current system support for low-bit KV cache struggles to deliver the expected speedup.} Previous implementations~\cite{kivi, atom, qserve} remain preliminary and case-specific, with significant room for further systematic optimization. A major bottleneck lies in the overhead introduced by quantization and dequantization. Although the KV cache is low-bit, the query (Q) values and attention scores remain in high precision. This results in mixed-precision matrix multiplications (mpGEMM), which existing hardware does not natively support, requiring dequantization before multiplication. Previous mpGEMM kernels like Ladder~\cite{ladder} and Marlin~\cite{marlin} are designed for low-bit weights but cannot be directly applied to low-bit KV caches. This is because weights are \textit{static and stored offline}, while KV caches are \textit{dynamic and generated online}. In autoregressive decoding, each newly generated token requires quantization, packing, and dequantization of the low-bit KV cache, introducing significant overhead and complexity in GPU kernel design, as illustrated in Fig.~\ref{fig:diff}.

\begin{figure}[t]
    \centering
    \includegraphics[width=0.95\linewidth]{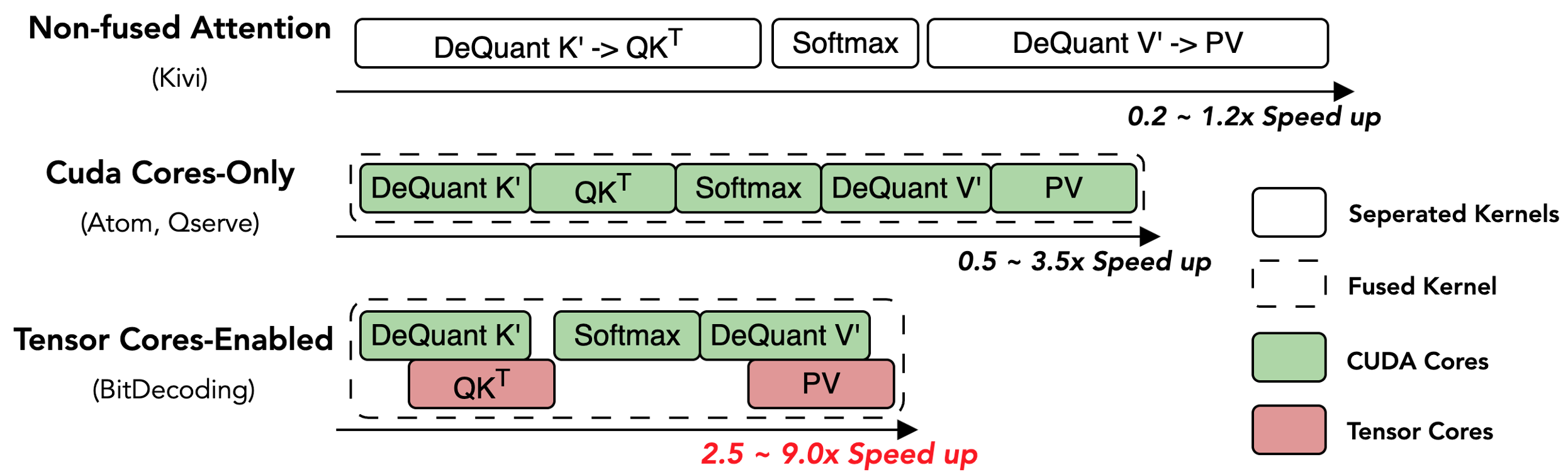}
    \caption{Comparison of different low-bit KV cache systems against half-precision FlashAttention. Each system follows the attention formulation \( \mathrm{Out} = \mathrm{softmax}(Q\,\mathcal{D}(K'^\top))\,\mathcal{D}(V') \), where \( K' \) and \( V' \) are low-bit quantized Key and Value tensors, and \( \mathcal{D}(\cdot) \) denotes the dequantization function.}
    \label{fig:compare}
\end{figure}

To address this, our insight is to leverage Tensor Cores for intensive matrix multiplications while efficiently utilizing CUDA cores for KV cache dequantization. Previous work either implemented with separated kernels or fused attention operations relied solely on CUDA cores, leaving Tensor Cores underutilized, as shown in Fig.~\ref{fig:compare}. Our approach is based on three key observations: First, modern language models employ Grouped-Query Attention (GQA) and Multi-Query Attention (MQA), which share a group of keys across multiple queries, enabling Tensor Cores to accelerate dot products in the self-attention mechanism. Second, leveraging Tensor Cores can alleviate computational pressure on CUDA cores, enabling more efficient execution of low-bit operations. Finally, newer GPU architectures provide distinct mechanisms: Hopper's support for asynchronous execution and warp specialization allows low-bit operations to overlap with computation~\cite{luo2024benchmarking}, while Blackwell's native support for low-precision formats (e.g., MXFP4) reduces these overheads by minimizing the need for on-the-fly data conversion.

Efficiently leveraging Tensor Cores for decoding with low-bit KV caches poses significant challenges. First, Tensor Cores require dequantized low-bit data to be aligned with high-precision formats, which is difficult in autoregressive decoding as the KV cache grows dynamically and must conform to Tensor Cores-specific layouts. Without optimized layouts, Tensor Cores may exhibit poor utilization or even produce incorrect results. Second, the high cost of dequantization can stall Tensor Cores execution, reducing GPU occupancy due to mismatched workloads between CUDA cores and Tensor Cores. Third, supporting low-bit KV caches across diverse attention mechanisms and quantization algorithms—with varying tensor-wise and channel-wise scaling—demands a general yet highly optimized implementation. Without careful design, either CUDA cores or Tensor Cores become performance bottlenecks during long-context generation.

To address the above challenges, we have designed and implemented \textbf{BitDecoding}, a high-performance long-context LLMs inference system with low-bit KV cache. The design of BitDecoding delivers several contributions essential for exploiting Tensor Cores, including: (i)~inducing low-bit optimized layouts based on hardware instructions, (ii)~aligning warps with residual buffer to saturate Tensor Cores, (iii)~re-mapping layouts for faster dequantization, and (iv)~coordinating kernels for quantization and dequantization. In addition, we contribute new strategies for parallelizing GPU warps to mitigate low-bit operations overhead, including (i)~efficient warp parallelism layout, and (ii)~enhancing attention algorithms for fast warp synchronization leveraging the GPU memory hierarchy.

We further contribute implementation techniques in BitDecoding for LLMs inference, including: (i)~a query transformation approach that enables efficient execution of diverse attention variants, allowing BitDecoding to be easily adopted in existing LLMs; (ii)~a high-performance quantization kernel that supports both channel-wise and tensor-wise scaling, ensuring generality across quantization algorithms; and (iii)~a dequantization kernel with a software-defined pipeline that coordinates CUDA and Tensor Cores for GEMM and dequantization, while overlapping data movement, including extra low-bit metadata; furthermore, BitDecoding incorporates architecture-specific optimizations that unlock Hopper’s warpgroup tensor operations and Blackwell’s native low-precision tensor formats to maximize decoding performance on the latest GPU generations.

BitDecoding is evaluated at both the kernel and end-to-end levels across Blackwell, Hopper, Ada, and Ampere GPU architectures. At the kernel level, it outperforms FP16 FlashDecoding-v2 by up to 8.6$\times$ on Blackwell (e.g., RTX 5090, using native MXFP4 format support), 8.0$\times$ on Hopper, 7.5$\times$ on Ada, and 4.8$\times$ on Ampere, while surpassing QServe by up to 4.3$\times$. At the end-to-end model level, BitDecoding reduces single-batch decoding latency by 3$\times$ on LLaMA-3.1-8B with a 128K sequence length and achieves over 4$\times$ higher serving throughput than QServe. 




\section{Background and Motivation}

\textbf{LLMs inference and low-bit KV cache.}
LLMs inference comprises two stages: (i) \emph{Prefill}, which processes the prompt and computes Key (K) and Value (V) tensors for caching; and (ii) \emph{Decode}, which updates the KV cache token-by-token for autoregressive generation. For a model with $n$ layers, $h_{kv}$ KV heads, and hidden size $d$, the KV cache requires $2 \cdot 16 \cdot n \cdot h_{kv} \cdot d \cdot b \cdot l$ bits (assuming FP16), where $b$ is the batch size and $l$ is the sequence length. Because this requirement grows linearly with both $b$ and $l$, the KV cache often dominates memory usage, especially for long-context and large-batch workloads. In batched inference, each sequence has an independent past context, so there is little batch-level parallelism or reuse when loading cached Keys and Values; \emph{consequently, KV-cache access is typically bound by memory bandwidth}. These constraints have spurred extensive research and industrial efforts on lower-bit KV caches~\cite{kivi, kvquant, 1bitkv} to reduce memory footprint and improve throughput while preserving accuracy close to non-quantized baselines.

\textbf{Tensor Cores and CUDA cores on modern GPUs.} When optimizing LLM inference and low-bit KV caches on GPUs, it is crucial to exploit both Tensor Cores and CUDA cores. Tensor Cores deliver the majority of compute FLOPS in modern GPUs but are specialized for matrix operations (e.g., GEMM), whereas CUDA cores provide more flexible vector, scalar, and control-flow capabilities at substantially lower peak FLOPS. For example, on the A100, Tensor Cores deliver up to 312 TFLOPS in FP16/BF16—far exceeding the 19.5 TFLOPS FP32 offered by CUDA cores. 

This performance gap has widened significantly in recent generations. The Hopper architecture introduces Warpgroup Matrix Multiply-Accumulate (WGMMA) instructions and warp-specialized pipelines to maximize asynchronous execution efficiency. The Blackwell architecture further exacerbates this disparity by supporting native micro-scaling formats (e.g., MXFP4, NVFP4), delivering up to 20 PFLOPS.

For fast LLM inference, substantial effort has gone into optimizing attention variants to exploit Tensor Cores. SOTA LLMs~\cite{liu2024deepseek, llama3, qwen3} increasingly adopt MQA~\cite{mqa} and GQA~\cite{gqa}, which reduce memory bandwidth by reusing KV heads across multiple queries. This reuse increases arithmetic intensity and improves compute efficiency~\cite{gta}, aligning well with the high-throughput, matrix-centric design of Tensor Cores. Consequently, leveraging Tensor Cores is becoming essential for efficient inference in long-context and grouped-attention LLMs.

\textbf{Limitations of existing low-bit KV cache systems.} To support low-bit KV caches for long-context LLM inference, a number of systems have been proposed~\cite{kivi,qserve,atom}. However, they often leave GPUs underutilized, leading to suboptimal performance. We summarize the key reasons below.

\begin{itemize}
\item \emph{Attention with separated low-bit KV-cache kernels:} The most straightforward approach, exemplified by Kivi~\cite{kivi}, decomposes mixed-precision attention into multiple standalone kernels and embeds them in a non-fused attention implementation. This design is highly flexible and readily supports many attention variants~\cite{gqa, mqa}. Yet the isolated launches repeatedly load and store intermediate data, inflate global-memory traffic, and break on-chip data reuse. The result is high launch overhead, increased memory bandwidth pressure, and lower effective throughput.

\item \emph{Fused attention with low-bit KV-cache kernels on CUDA cores solely:} Given the generality of CUDA cores for mixed-precision operations, a natural extension of FlashAttention-style fusion~\cite{dao2023flashattention2} is a CUDA-cores–only implementation of low-bit KV caches. While this outperforms non-fused designs, it still underutilizes Tensor Cores. In these systems, both dequantization and matrix operations (GEMV/GEMM) are executed on CUDA cores via fused multiply–add (FMA) instructions. Under mixed precision, CUDA cores must handle expensive dequantization (e.g., int4/8 $\rightarrow$ FP16/BF16), scaling, and element-wise ops—tasks that are memory-bound and consume instruction slots, register bandwidth, and L1/L2 capacity. This reduces occupancy and limits tile sizes, leaving fewer resources for the compute-heavy matrix multiplications. Consequently, running both dequantization and matmul on CUDA cores introduces significant overhead, especially for attention variants with higher arithmetic intensity. 
\end{itemize}

\section{Proposed Solutions and Challenges}


\subsection{Solution: Cooperative use of Tensor Cores \& CUDA Cores} \label{sec:opportunity}

In this paper, we want to explore a solution that can achieve a \emph{cooperative} use of Tensor Cores and CUDA cores to support low-bit KV caches during long-context LLMs inference. Our design introduces new designs and implementations that (i) construct and schedule matrix multiplications on Tensor Cores, and (ii) execute non-matrix-multiplication operations—quantization, packing and dequantization—efficiently on CUDA cores. To make this cooperation effective, we balance workloads across the Tensor Cores and CUDA cores and carefully orchestrate data movement so that dequantization feeds Tensor-Core GEMM without stalls, memory traffic is minimized, and end-to-end decoding throughput is maximized.

To ensure broad adoption, we aim to realize this cooperative design as a system that (i) supports low-bit KV caches across multiple attention variants (including MHA, MQA, and GQA), and (ii) spans multiple GPU generations. The former requires a clean interface that integrates with existing attention implementations; the latter requires designs that are easy to adapt, enabling rapid targeting of different GPU backends while sustaining high decoding throughput.

We expect significant benefits from this proposed solution. For example, by enabling low-bit decoding that builds on FlashAttention-3 (FA-3)~\cite{fa3}, we can leverage SM90-specific features—such as warp-specialized pipelines—that yield up to $6\times$ speedups over prior implementations, avoiding the 35\% throughput penalty associated with legacy SM80 instructions. Furthermore, this design anticipates the architectural capabilities of Blackwell, where native support for low-precision formats will drive even more substantial throughput improvements.

\subsection{Open challenges}
Although promising, the \emph{cooperative} use of Tensor Cores and CUDA cores for low-bit KV caches is particularly challenging to implement for several reasons:

\begin{figure}[t]
    \centering
    \begin{subfigure}[b]{0.492\linewidth}
        \centering
        \includegraphics[width=\linewidth]{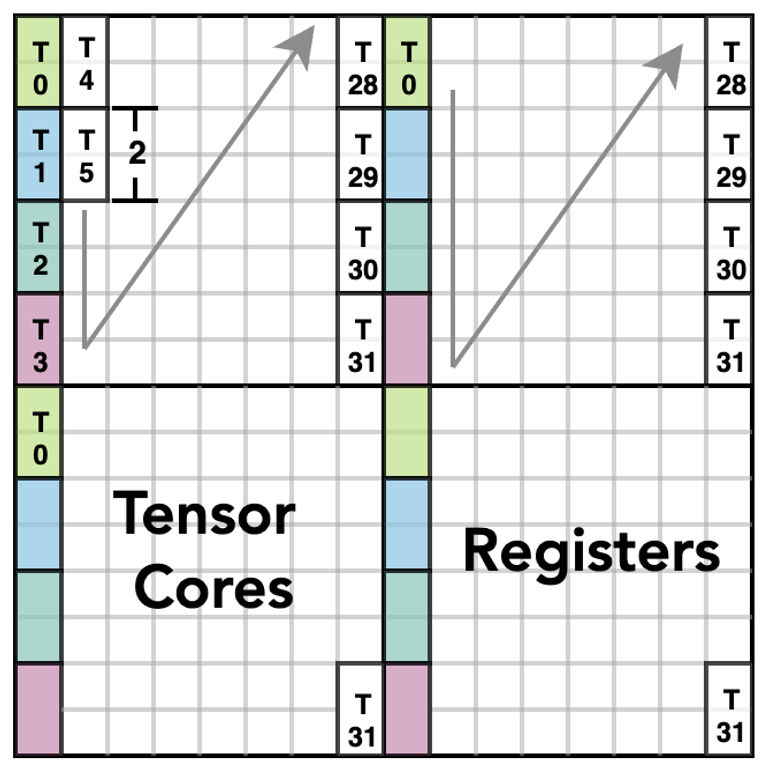}
        \caption{FP16 Fragment layout}
        \label{fig:ldmatrix}
    \end{subfigure}
    \hfill
    \begin{subfigure}[b]{0.492\linewidth}
        \centering
        {\includegraphics[width=\linewidth]{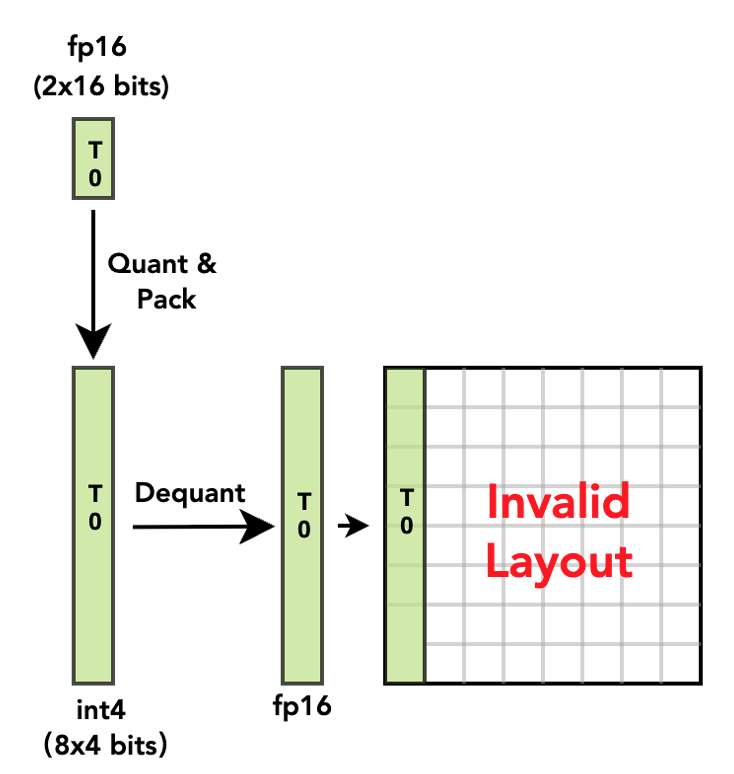}}
        \caption{Int4 Fragment layout}
        \label{fig:invalid}
    \end{subfigure}
    \caption{(a) \texttt{mma.m16n8k16} fragment layout for matrix B. Each thread ($T_i$) is assigned a specific set of values based on the instruction-defined interleaved mapping. (b) For INT4, quantization packs values contiguously per thread. After dequantization, the layout misaligns with the expected interleaved pattern.}
    \label{fig:mismatch}
\end{figure}

\textbf{Challenge 1: Tensor Cores often suffer from low-bit layout mismatches.} \label{sec:mismatch} Aligning low-bit data layouts with Tensor Cores requirements is difficult, especially in autoregressive generation where KV caches expand dynamically. 

At runtime, after quantization and packing, the low-bit KV cache must dequantize into a half-precision layout that matches what Tensor Cores expect. This matching is challenging for three reasons. 

First, fragment layouts vary across instructions and GPU generations. After using the optimized data-movement instruction \texttt{ldmatrix}, the fragment residing in registers enforces a strict value-to-thread mapping. Fig.~\ref{fig:ldmatrix} illustrates the registers read by each thread ($T$) for \texttt{mma.m16n8k16} with repeat tiling along the $N$ dimension. However, this mapping differs from other Tensor Core instructions (e.g., \texttt{mma.m16n8k8}) and from Hopper’s \texttt{wgmma} family (e.g., \texttt{wgmma.m64n64k16}).

Second, low-precision bitwidths exacerbate alignment issues. Although Tensor Cores instructions require specific compute types, their rigid, interleaved register layout makes lower-precision data hard to match directly. Without a layout transform, the low-bit register layout becomes an \textbf{invalid layout} for MMA execution due to misalignment with the interleaved access patterns. As shown in Fig.~\ref{fig:invalid}, two FP16 values originally computed by Thread 0 (T0) may be quantized and packed as eight consecutive low-bit values in the KV cache; after unpacking and dequantization, they no longer align with the expected Tensor Core register layout, yielding incorrect values. Even with native low-precision formats in Blackwell, hardware support remains limited, especially for the KV cache, which still depends on continuous quantization and packing; software must therefore carefully handle low-precision values and micro-scaling factors~\cite{nvidia_triton_blackwell_2025}.

Finally, dequantization can bottleneck execution: naive low-bit$\rightarrow$FP16 casts are slow~\cite{lop3} and require a \textbf{friendly layout} to run efficiently. Prior work such as Ladder~\cite{ladder} and Marlin~\cite{marlin} mitigates mismatch for static weights by inserting separate layout-transformation kernels, but this adds substantial overhead and is unsuitable for dynamic decoding. Experimental details are given in Table~\ref{tab:quantization}.

\begin{figure}[t]
    \centering
    \begin{subfigure}[b]{0.492\linewidth}
        \centering
        \includegraphics[width=\linewidth]{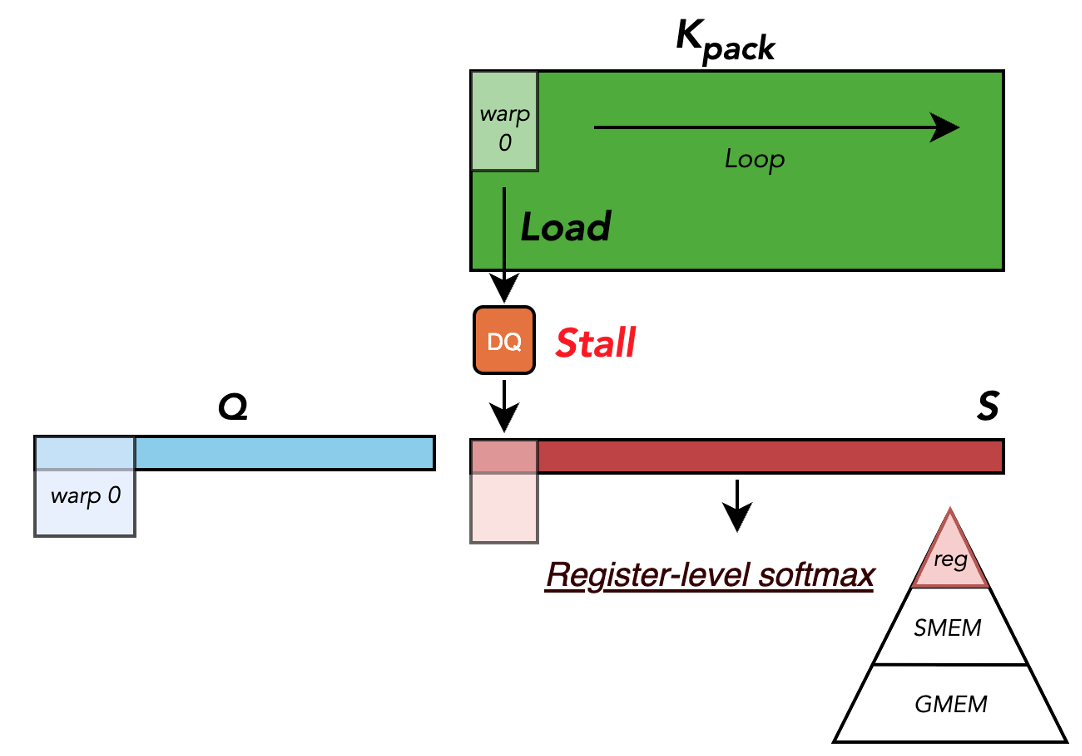}
        \caption{Original Warp Design}
        \label{fig:single_warp}
    \end{subfigure}
    \hfill
    \begin{subfigure}[b]{0.492\linewidth}
        \centering
        {\includegraphics[width=\linewidth]{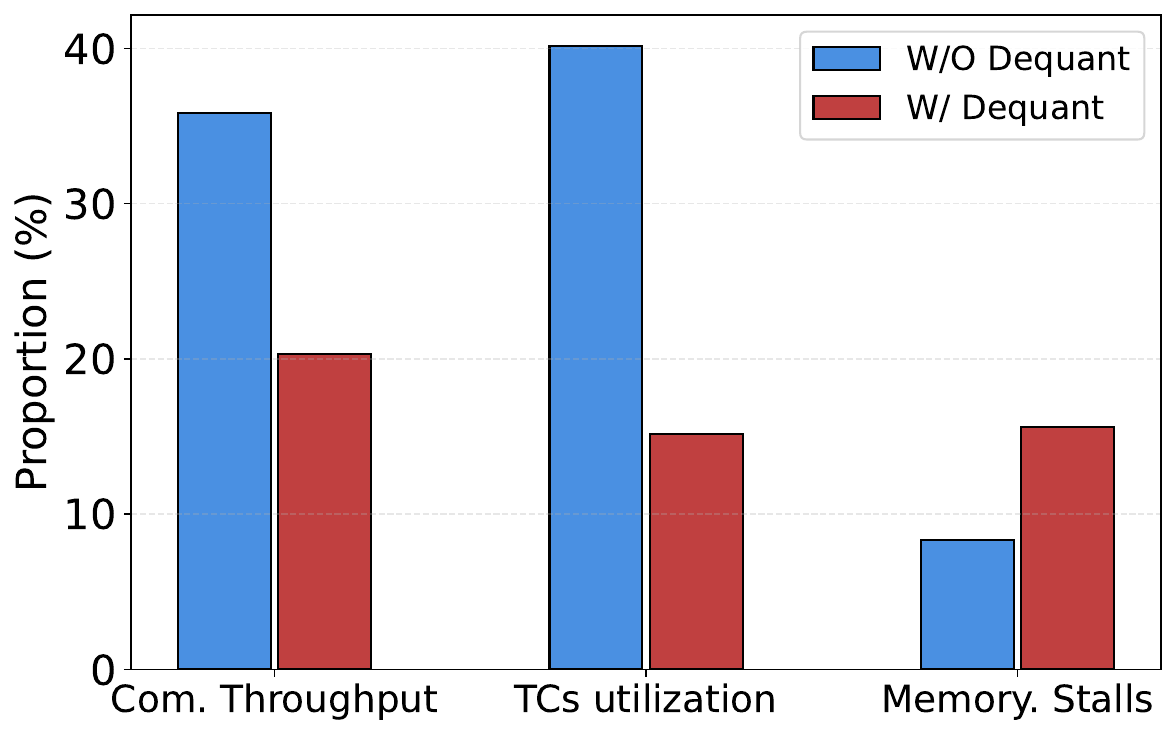}}
        \caption{Micro-level Analysis}
        \label{fig:with_dequant}
    \end{subfigure}
    \caption{(a) A single warp along $N$ for register-level operations will experience stalls due to dequantization (DQ) (b) Micro-level comparision with and without dequantization.}
    \label{fig:warp_layout}
\end{figure}

\textbf{Challenge 2: Frequent stalls limit Tensor Cores utilization.} We observe that empirically tuned warp layouts and partitioning in high-performance attention kernels often inadvertently degrade low-bit KV-cache performance.

Under FlashAttention’s original warp partitioning, the additional dequantization (DQ) can substantially reduce throughput and Tensor Core utilization. As shown in Fig. \ref{fig:single_warp}, FlashAttention assigns a single warp along the $N$ dimension to perform register-level softmax and the matrix multiplication $PV$, with $P$ stored in registers aligned to the Tensor Core layout. When DQ is inserted before the matmul, this strategy becomes inefficient: small warp tiles of $K$ or $V$ must traverse $N$ sequentially, so DQ frequently stalls the warp. Nsight Compute profiling\cite{NVIDIA_Nsight_Compute} in Fig. \ref{fig:with_dequant} confirms that the added DQ overhead increases memory-access stalls and depresses compute throughput and Tensor Cores utilization, consistent with prior observations\cite{fan2025warpdrive}.

Furthermore, native low-precision formats introduce their own overhead despite eliminating dequantization. Specifically, to utilize low-precision Tensor Cores for the second matrix multiplication ($PV$), the probability matrix $P$ must be dynamically re-quantized after the softmax operation: $P_{f16} = \mathrm{softmax}(Q_{f4}K_{f4}^T), \quad O_{f16} = \mathbf{Quant}(P_{f16})V_{f4}$. This on-the-fly quantization creates a new computational bottleneck that can similarly stall Tensor Cores execution.

\textbf{Challenge 3: Lack of generalizable system optimizations for different low-bit KV-cache methods.} Popular KV-cache quantization methods use diverse scaling granularities for the Key tensor—tensor-wise~\cite{atom,kvquant} and channel-wise~\cite{kivi,gear}—which complicates building a unified system that supports them all. Online quantization and packing require reductions and element-wise transforms, adding nontrivial runtime overhead. Moreover, auxiliary metadata (scale and zero-point) increases memory traffic and, without careful scheduling, disrupts the load–compute pipeline. Prior mixed-precision kernel optimizations~\cite{ladder,marlin} target static weight quantization and do not generalize to the dynamic, step-by-step nature of KV caches. To date, generalizable system-level optimization techniques for high-performance, low-bit KV-cache quantization are lacking.






\section{BitDecoding Design} \label{sec:design}

In this section, we present the design of BitDecoding system which realizes the cooperative use of Tensor Cores and CUDA cores in supporting low-bit KV cache. The design primarily contains (i) new methods and principles for optimizing the low-bit layout in using Tensor Cores, and (ii) new strategies for parallelizing and coordinating GPU warps that can minimize the stalls due to dequantization.

\begin{figure}[t]
    \centering
    \includegraphics[width=0.9\linewidth]{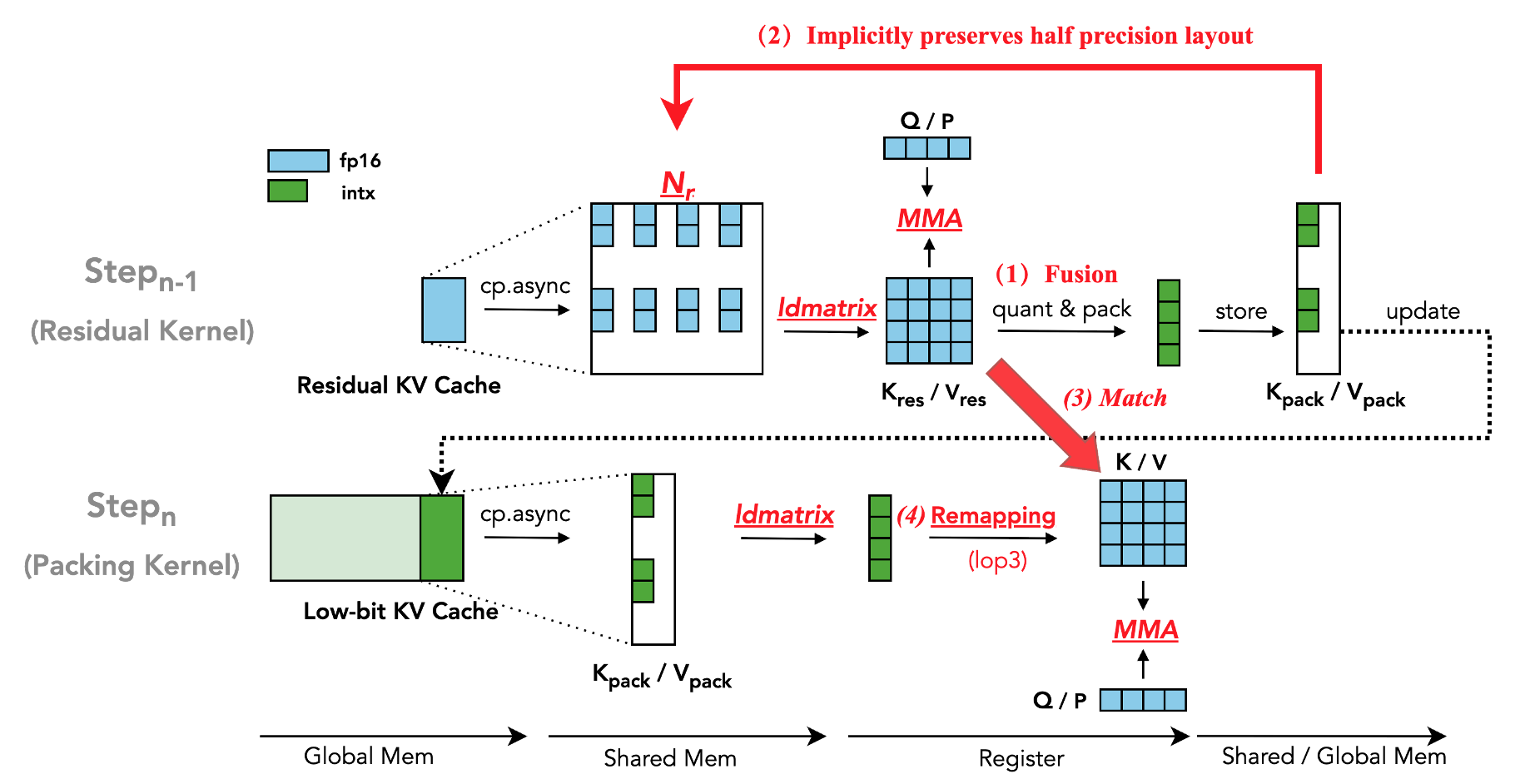}
    \caption{Overview of methods for optimizing low-bit layout on Tensor Cores. (1) Fused computation and quantization within Tensor Cores fragments. (2) The low-bit packing data preserves FP16 values. (3) Low-bit Layout matches with the dequantized half-precision layout. (4) Layout remapping for faster dequantization.}
    \label{fig:scheme}
\end{figure}

\subsection{Methods for optimizing low-bit layout on Tensor Cores} \label{subsec:Scheme}

The first challenge our design aims to address is to ensure BitDecoding can automatically generate an optimized layout that can fully utilize Tensor Cores across different GPU generations and different configurations of the low-bit KV caches. For this, we have designed the following principles and methods:

\textbf{(1) Inducing low-bit optimized layout with hardware instructions.} Our design is motivated by a novel insight: the thread-to-register mapping of \texttt{ldmatrix} loads data in Tensor Core’s interleaved fragment layout. As shown in Fig.~\ref{fig:scheme}-(2), if each thread then quantizes and packs locally, the resulting low-bit packing \emph{implicitly preserves} the half-precision (FP16) interleaved layout. On unpacking and dequantization, values already match Tensor Core registers—no global reshape is required. Thus, rather than relying on heavyweight global transforms via manual implementations~\cite{marlin} or iterative search~\cite{ladder} as in prior methods, we use hardware instructions to automatically induce a valid low-bit packing layout while computing. This yields zero-overhead remapping that is efficient, compatible with Tensor Cores execution, and avoids extra data movement.

Building on this insight, we design a dedicated GPU \textit{Residual Kernel} that fuses computation, quantization, and packing for newly generated FP16 KV tensors. Using \texttt{ldmatrix}, we load the high-precision KV tensor into registers structured for Tensor Cores, perform the matrix operation (e.g., $QK^\top$ or $PV$), and then have each thread quantize and pack its portion in registers (see Fig.~\ref{fig:scheme}-(1)). The result is interleaved, layout-compatible low-bit data written directly to global memory, updating the low-bit KV cache.

To consume this cache, we introduce a \textit{Packing Kernel} that fuses dequantization with computation. To guarantee correct register layout during unpacking, it mirrors the Residual Kernel’s instruction configuration which (i)~uses the same \texttt{ldmatrix} variant and (ii)~follows the same \texttt{mma} variant and warp-tiling configuration. Consequently, when the Packing Kernel loads packed low-bit data via \texttt{ldmatrix}, the unpacked values are inherently aligned with Tensor Core registers and can participate in matrix multiplication immediately, without explicit layout correction.

\textbf{(2) Aligning warps with residual KV cache to saturate Tensor Cores.} Tensor Cores execute warp-tiled matrix operations, which require input tiles to be fully populated to achieve optimal throughput. Based on this, \emph{our insight} is that by allocating a residual buffer with size matching the tiling capacity of Tensor Cores, we ensure that low-bit data aligns with the compute granularity of the hardware to fully utilize the computing ability of the computing unit. 

To implement this idea, we introduce a half-precision residual KV cache with a residual block size $N_r$. Let $X \in \mathbb{R}^{L \times d}$ denote the entire KV cache. We partition $X$ into:
\[
X = X_{\text{pack}} \cup X_{\text{res}}, \quad \text{where} \quad
\begin{cases}
X_{\text{pack}} = X[:L - N_r] \\
X_{\text{res}} = X[L - N_r:]
\end{cases}
\]

We define $\beta$ as the bit-width for low-bit quantization (e.g., $\beta = 4$ or $2$), and $\omega$ as the word size used for packed storage (e.g., $\omega = 16$ for INT16). The corresponding \textit{packing ratio} is given by $R = \omega / \beta$. Let $W_n$ denote the number of warps along the N dimension, and $P_n$ the number of elements each warp tile processes (e.g., $P_n = 8$ under \texttt{mma.m16n8k16}). To ensure each Tensor Cores fragment is fully populated for each warp, the residual block size is computed as:
\begin{equation}
N_r = P_n \times W_n \times R
\label{eq:residual}
\end{equation}
This guarantees that low-bit KV cache fragments align precisely with the warp-level tiling of Tensor Core operations, enabling dense, layout-compatible packing and maximizing compute unit occupancy.



\textbf{(3)~Re-mapping layout for faster dequantization.} Though compatible with Tensor Cores layout, the layout is inefficient to dequantization due to directly casting low-bit values to FP16 using \texttt{static\_cast} introduces significant overhead. 

To mitigate this inefficiency, we further design a faster dequantization mapping approach based on low-level bitwise operations and instructions inspired by~\cite{lop3}. After loading packed data into registers using \texttt{ldmatrix}, we cast them to INT32 before mapping them to the interleaved Tensor Core layout following the 75316420 pattern. This layout enables efficient conversion of INT4/INT2 data to FP16 using the \texttt{lop3} instruction for bitwise manipulation while aligning with the Tensor Core computation pattern.

\textbf{(4)~Coordinating Residual and Packing Kernels with Configuration Setup.} 
This design is executed by coordinating the Residual and Packing kernels under a unified instruction configuration. First, the hardware instruction configuration—including \texttt{ldmatrix} and \texttt{mma} variants—can be determined based on GPU architectures. With this configuration, the residual block size $N_r$ is computed based on the bit-width of the low-bit KV cache. As shown in Fig.~\ref{fig:scheme}, the Residual kernel loads high-precision KV entries into registers via \texttt{ldmatrix}, performs computation using Tensor Cores, and then fuses quantization and packing before storing the results into the low-bit KV cache. The Packing kernel, using the same instruction configuration, loads the packed data into registers, performs efficient dequantization, and proceeds with Tensor Core computation.


\subsection{Strategies for parallelizing warps} \label{subsec:Warp}

The second challenge is ensuring BitDecoding avoids the pitfalls of existing warp-parallelization strategies for mixed-precision attention, which suffer from low hardware utilization due to frequent warp stalls. Our key insight is that low-bit data moves at much higher bandwidth than full precision, shifting the bottleneck from memory to compute. We therefore design a warp layout that exploits the GPU memory hierarchy to parallelize low-precision operations efficiently, minimizing data movement and substantially improving Tensor Cores utilization (Table~\ref{tab:warp_softmax} demonstrates minimal overhead).

\begin{figure}[t]
    \centering
    \includegraphics[width=0.71\linewidth]{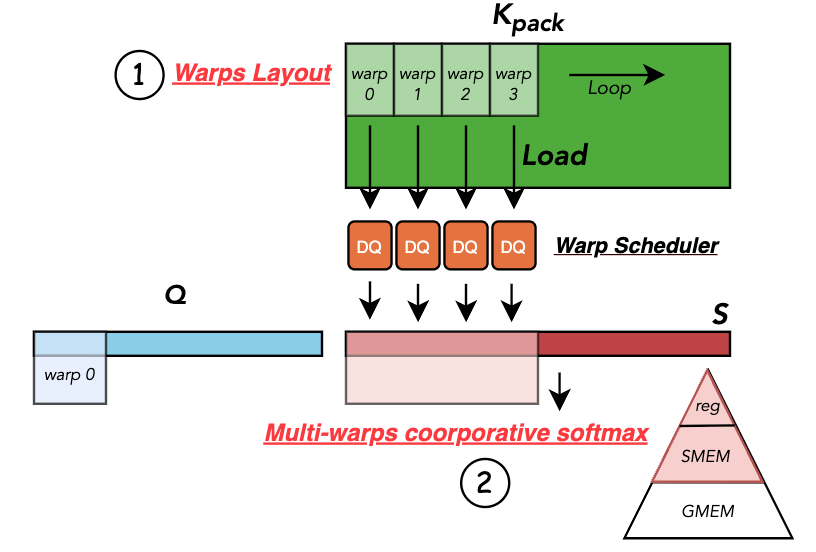}
    \caption{Enhancing parallism for efficient Tensor Cores utilization with (1) new warp layout design reduces dequantization stalls and (2) cooperative softmax leverages data movement between GPU register and shared memory for cross-warp reduction with minimal overhead.}
    \label{fig:multi_warps}
\end{figure}

\textbf{(1)~Enhancing warps parallelism for low-precision operations.} We introduce a novel warps layout to enable parallel operations of multiple packed data chunks. Using dequantization as an example, we modify the warp partitioning strategy to better exploit parallelism. As illustrated in Fig.~\ref{fig:multi_warps}, instead of the original strategy that allocates multiple warps along the $M$ dimension, we constrain the allocation to $W_m = 1$—leveraging the fact that the decoding query length is typically small ($<16$)—and reallocate resources to increase the number of warps along the $N$ dimension ($W_n$).

By increasing $W_n$, dequantization stalls can be effectively mitigated by the Streaming Multiprocessor (SM) warp scheduler~\cite{warp_schedule}, as multiple warps concurrently execute dequantization on packed data before proceeding to Tensor Cores-based matrix multiplication.

Similarly, this parallelism strategy alleviates the stalls introduced by on-the-fly quantization in native low-precision attention, ensuring that neither quantization nor dequantization becomes a serialization bottleneck.

\textbf{(2)~Leveraging memory hierarchy for warps synchronization.} However, with results now distributed across different registers and warps, the original register-level softmax becomes infeasible. Moreover, \emph{a key challenge emerges} due to the incompatibility between the new warp layout and the expected format for MMA operations on $PV$. 

To address this, we leverage a multi-level memory hierarchy—spanning registers and shared memory—to enable cross-warp reduction and synchronization for the softmax computation. As illustrated in Algorithm~\ref{alg:warp}, we extend existing high-performance attention algorithms, such as FlashAttention, by introducing two additional shared memory buffers: \( sTMP \in \mathbb{R}^{W_n} \) and \( sAcc \in \mathbb{R}^{T_m \times T_n} \). The buffer \( sTMP \) facilitates cross-warp reduction for computing the row-wise maximum during softmax. This is achieved by first performing intra-warp reduction within registers, followed by inter-warp reduction via shared memory. The buffer \( sAcc \) temporarily stores the attention scores \( P \) computed in Tensor Core registers and later reloads them via \texttt{ldmatrix}, ensuring proper alignment for subsequent Tensor Core \texttt{mma} operations.

Since $W_n$ is typically small, we reuse the shared memory pointer of $sTMP$ for $sAcc$ to minimize memory overhead. Moreover, on Hopper Tensor Cores, WGMMA supports direct shared memory access, eliminating the need for explicit data movement from shared memory to registers.

\begin{algorithm}
\small
\caption{Multi-warps Cooperative Softmax} \label{alg:warp}
\begin{algorithmic}[1]
\Require \( sTMP \in \mathbb{R}^{W_n}\) and \( sAcc \in \mathbb{R}^{T_m \times T_n} \) in SMEM.
\Require Load \( Q_i \in \mathbb{R}^{T_m \times d} \) and \( K_i, V_i \in \mathbb{R}^{T_n \times d} \) to REG.
\State \( S_i = Q_i K_j^T \) where \( S_i \in \mathbb{R}^{T_m \times T_n} \).
\State \( m_i^{new} = \max(m_i, \textcolor{commentcolor}{\text{rowmax}(S_i, sTMP)}) \).
\State \( P_i = \exp(S_i - m_i^{new}) \) where \( P_i \in \mathbb{R}^{T_m \times T_n} \).
\State $sAcc = \textcolor{commentcolor}{\text{tiled\_copy\_r2s}(P_i)}$.
\State $P_i' = \textcolor{commentcolor}{\text{tiled\_copy\_s2r}(sAcc)}$
\State \( O_i^{new} = P_i' V_j + \text{diag}(e^{m_i - m_i^{new}}) O_i \).
        
\end{algorithmic}
\end{algorithm}





\begin{figure*}[t]
    \centering
    \includegraphics[width=0.95\linewidth]{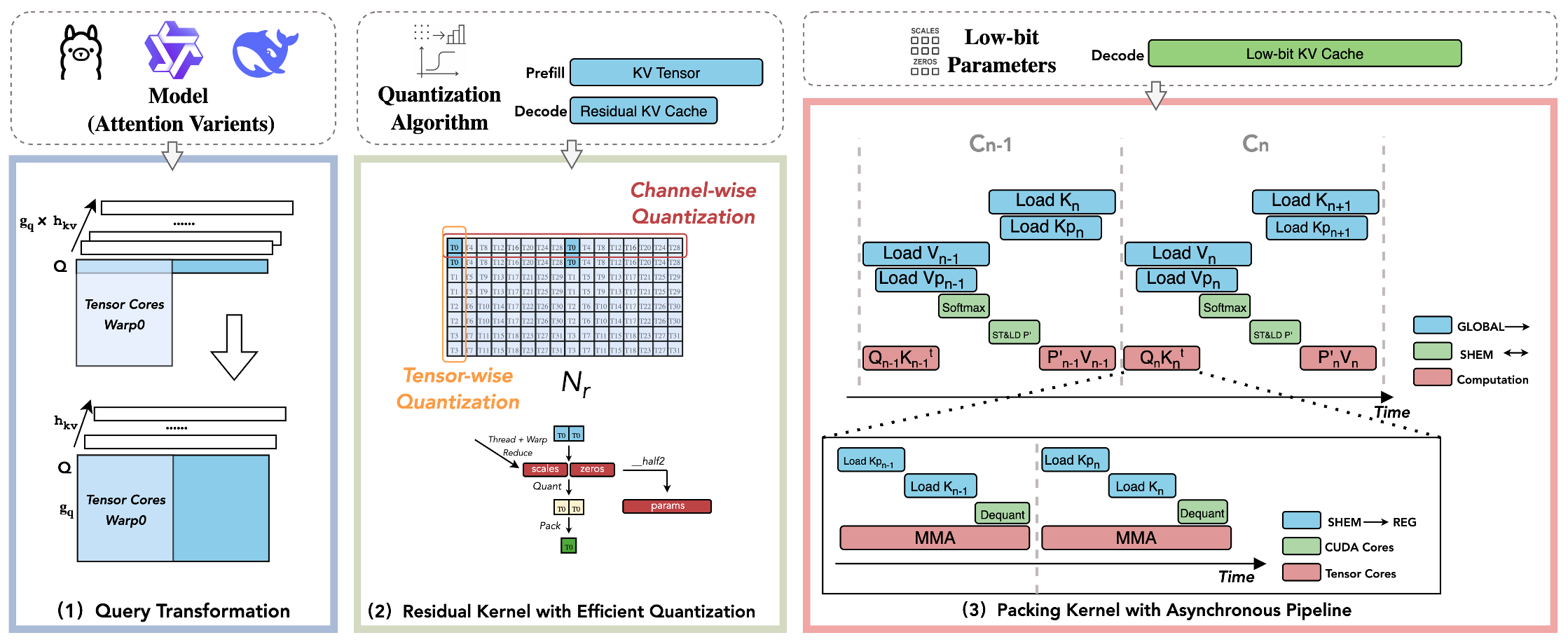}
    \caption{System overview of BitDecoding. 
    (1) \textbf{Query Transformation} restructures the query tensor layout to enable efficient warp-level execution for attention variants on Tensor Cores. 
    (2) \textbf{Residual Kernel} performs quantization and packing with minimal overhead, supporting both tensor-wise and channel-wise scaling. 
    (3) \textbf{Packing Kernel} executes dequantization and matrix multiplication using a fine-grained, asynchronous pipeline, maximizing Tensor Cores and CUDA Cores utilization with low-bit parameters.}
    \label{fig:overview}
\end{figure*}

\section{System Implementation}

In this section, we describe how we implement BitDecoding, as illustrated in Fig.~\ref{fig:overview}. Our implementation consists of three major components: (i) a \emph{query transformation} component that supports diverse attention variants in LLMs; (ii) a \emph{Residual Kernel} that performs low-cost quantization and packing while remaining general to both tensor-wise and channel-wise scaling across quantization algorithms; and (iii) a \emph{Packing Kernel} with a fine-grained pipeline that fully utilizes both Tensor Cores and CUDA cores. Finally, we discuss architecture-specific optimizations that leverage the advanced features of the latest GPU generations (e.g., Hopper and Blackwell) to further enhance decoding throughput.

\subsection{Query Transformation} \label{sec:query}

Modern LLMs adopt diverse attention variants~\cite{llama3,qwen3,liu2024deepseek} with different key–value (KV) sharing patterns. BitDecoding aims to support all these variants. 

For instance, in GQA and MQA, multiple query heads share a KV head, reducing the number of KV projections and memory accesses. The degree of sharing is measured by $g_q = h_q / h_{kv}$, where $h_q$ and $h_{kv}$ are the numbers of query and KV heads, respectively: $g_q = 1$ corresponds to MHA, $g_q > 1$ denotes GQA, and $h_{kv}=1$ (i.e., $g_q = h_q$) characterizes MQA.

A challenge arises in decoding: since $Q\_len = 1$ (one token at a time), the query tensor has a very small batch dimension, and a naive $Q \cdot K^{\top}$ underfills Tensor Cores, yielding poor warp occupancy and low throughput.

To address this, we perform a \emph{query transformation} that reorganizes the query layout to better match Tensor Core tiling. As illustrated in Fig.~\ref{fig:overview} (left), we reshape the query tensor from $[1,(g_q,h_{kv})]$ to $[g_q,h_{kv}]$, effectively forming a larger $Q$ tile without changing the semantics of attention or its KV-sharing pattern. Grouped query heads are then processed in parallel as a larger GEMM block, fully populating Tensor Core fragments, improving warp occupancy, and increasing throughput.

\subsection{Residual Kernel} 

A primary challenge in low-bit KV-cache design is supporting diverse quantization algorithms—especially differing scaling granularities (e.g., tensor-wise, channel-wise)—without sacrificing performance. Quantization involves reductions and element-wise operations to compute scale and zero-point, followed by bit-packing; during decoding these must run online, adding runtime overhead and risking misalignment with the rigid layouts expected by Tensor Cores. To address this, we design the \emph{Residual Kernel} with two key optimizations:

\textbf{(1) Partitioning KV cache based on residual block size.}
During prefill with context length $L$, we split the KV cache based on a Tensor Cores-aligned residual block size $N_r$ (see Eq.~\ref{eq:residual}). The first $N_p = L - (L \mod N_r)$ entries are quantized and packed into the low-bit KV cache using a fused quantization and packing operation. The remaining KV Tensor with size $\texttt{res\_len} = L \mod N_r$ are stored in the half-precision residual KV cache. At each decode step, the newly generated $K, V$ tensors are appended to the residual cache and used for attention computation. This cache grows incrementally until it reaches the residual block size $N_r$. Once per token generation, the Residual Kernel computes attention using the half-precision residual KV cache and optionally quantizes it (when $\texttt{res\_len} = N_r$) into packed format.

With this KV cache partitioning during decoding, we can naturally perform channel-wise quantization along the $seq\_len$ and tensor-wise quantization along the hidden dimension within the residual block.

\textbf{(2) Optimizing reduction with warp-level instructions.} As shown in Fig.~\ref{fig:overview} (mid), once the half-precision KV data is computed, it remains in registers as Tensor Cores fragments—structured in the native interleaved layout used by \texttt{mma} operations. To efficiently compute the quantization parameters (scale and zero-point), we first perform thread-level reductions to obtain local min/max statistics within each group.

These local results are then aggregated across the warp using the PTX instruction \texttt{\_\_shfl\_xor\_sync}, enabling efficient warp-level reduction without shared memory. When the warp repetition factor $W_n > 1$, we introduce a small shared memory buffer to coordinate the final reduction across warps.

After computing the quantization parameters, each thread performs in-register quantization and packs the low-bit values into INT16 format. This avoids extra memory movement and keeps data in a computation-ready state. To minimize overhead, both the scale and zero-point are stored in a compact \texttt{half2} format, enabling efficient memory access and fused multiply-add during dequantization in the decode phase.

\subsection{Packing Kernel} 

Another challenge is the auxiliary low-bit metadata (scale and zero-point), which increases memory traffic, while dequantization still runs on CUDA cores. Without careful scheduling, this disrupts the load–compute pipeline and prevents overlap with Tensor Core operations. We therefore design a fine-grained asynchronous pipeline: CUDA cores handle dequantization, Tensor Cores execute matrix multiplications, and both are orchestrated to overlap with memory transfers through the GPU hierarchy—enabling efficient mixed-precision computation.

\textbf{(1) Optimizing asynchronous data movement.} \emph{From Global to Shared Memory}, we follow FlashAttention~\cite{dao2023flashattention2} via block-wise tiling~\cite{wang2025tilelang} and strategic recomputation. It processes input matrices \( Q \in \mathbb{R}^{T_m \times d} \), \( K, V \in \mathbb{R}^{T_n \times d} \) in tiles within shared memory, using block sizes \( T_m \) and \( T_n \). The number of key-value tiles is \( C_n = \lceil L / T_n \rceil \).

To efficiently manage quantization parameters, we introduce dedicated shared memory buffers for quantization paramenter $K_{pack}$ params ($K_p$) and $V_{pack}$ param ($V_p$), facilitating efficient tiling for memory copy. These buffers store \texttt{scale} and \texttt{zeros} in the \texttt{half2} format, allowing them to be loaded in a single instruction.

The shape of $K_p$ is determined by the quantization granularity setting, and the $V_p$ follows a Tensor-wise layout:
\begin{itemize}
    \item \textbf{Channel-wise:} $(T_n / \text{group\_size}, d)$.
    \item \textbf{Tensor-wise:} $(T_n, d / \text{group\_size})$.
\end{itemize}

To achieve optimal memory overlapping, all global-to-shared memory transfers are executed asynchronously using the \texttt{cp.async} intrinsic, ensuring efficient pipeline execution, as shown in Fig.~\ref{fig:overview} (right). We optimize memory transactions using instructions with different caching strategies:

\begin{itemize}
    \item \textbf{\texttt{cp.async.cg}}: Used for $Q$, $K_{\text{pack}}$, and $V_{\text{pack}}$, which cache only in global memory as they are not reused within the same kernel.
    \item \textbf{\texttt{cp.async.ca}}: Applied to $K_p$ and $V_p$, ensuring smaller byte-level alignment for fine-grained memory access.
\end{itemize}

In Hopper architecture, we follow FA3, leveraging the \texttt{tma.copy} instruction for data loading. This facilitates warp-specialized scheduling, improving data locality and reducing memory latency across multiple warps.

\emph{From Shared Memory to Register}, we use the PTX instruction \texttt{ldmatrix} to efficiently load $K_{\text{pack}}$, $V_{\text{pack}}$ and $sAcc$ from shared memory into registers with the Tensor Cores tiling layout. To eliminate bank conflicts, we use a sizzling scheme~\cite{cutlass2024} defined as:
\begin{equation} 
\text{col}_{id} = \text{row}_{id} \oplus \text{col}_{id} 
\end{equation}
achieve bank conflict-free access. Additionally, we restructure the shared memory layout of $K_p$ and $V_p$ to further reduce bank conflict and maximize throughput efficiency.

\textbf{(2) Asynchronous pipeline for overlapping CUDA Cores and Tensor Cores.}
To fully utilize both CUDA cores and Tensor Cores, we implement a register-level, asynchronous pipeline that overlaps computation with memory operations. In this pipeline, shared-memory loads via \texttt{ldmatrix} and dequantization (\texttt{Dequant}) run concurrently with Tensor Core matrix multiplications (\texttt{mma}) under the SM warp scheduler.

As shown in Fig.~\ref{fig:overview} (right), while the $i$-th slice is being processed by \texttt{mma} on Tensor Cores, the $(i+1)$-th slice is simultaneously loaded from shared memory (\texttt{ldmatrix}) and dequantized. This sustains a continuous producer–consumer flow, improving instruction throughput and maximizing utilization of both CUDA cores and Tensor Cores.
\begin{figure*}[t]
    \centering
    \begin{subfigure}[b]{0.49\linewidth}
        \centering
        \includegraphics[width=\linewidth]{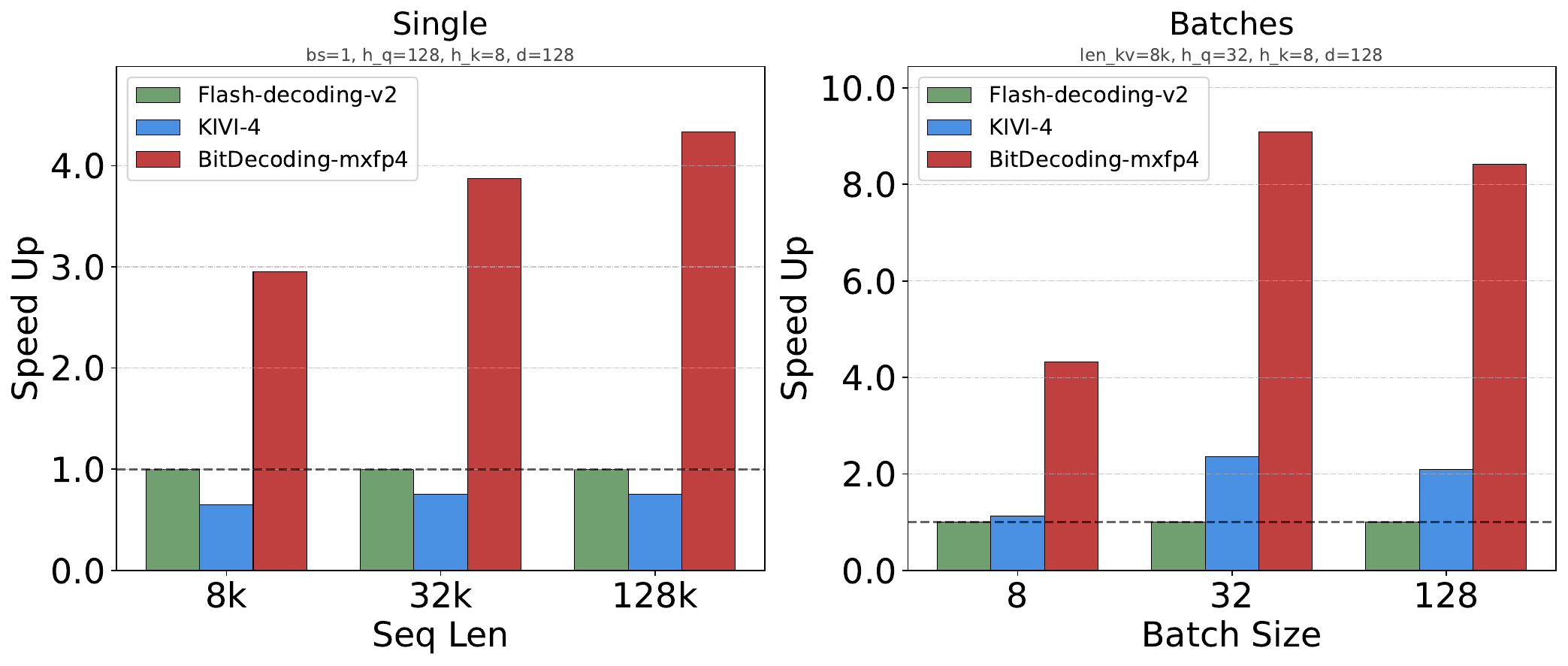}
        \caption{RTX 5090}
        \label{fig:kernel_5090}
    \end{subfigure}
    \hfill
    \begin{subfigure}[b]{0.49\linewidth}
        \centering
        \includegraphics[width=\linewidth]{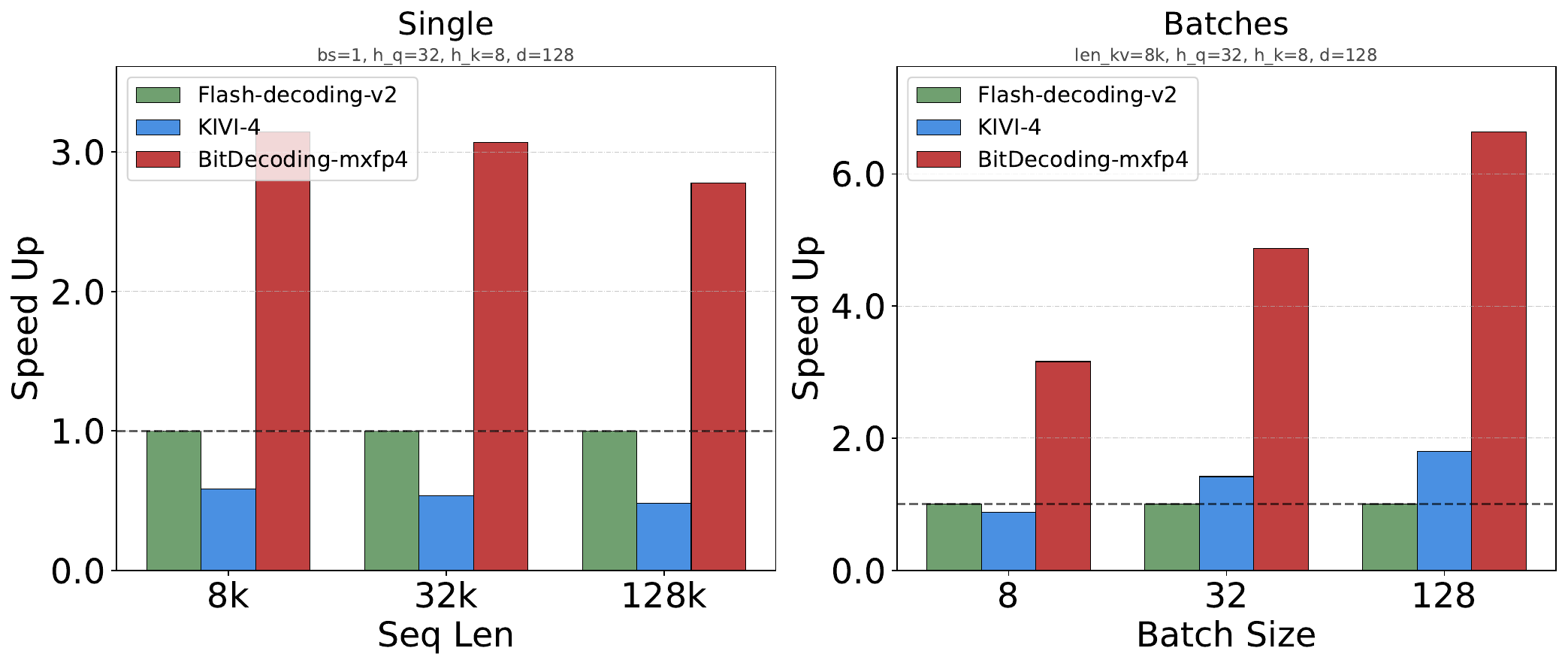}
        \caption{RTX PRO 6000}
        \label{fig:kernel_6000}
    \end{subfigure}
    \caption{Kernel performance with mxfp4 on Blackwell architectures.}
    \label{fig:blackwell}
\end{figure*}

\subsection{Latest Architectures Support} 
While the design presented thus far effectively targets pre-Hopper architectures (e.g., Ampere), newer generations introduce distinct hardware features that require tailored optimization strategies. Below, we detail how our approach adapts to leverage the specialized instructions and native data formats of the Hopper and Blackwell architectures.

\textbf{(1) Unlocking Hopper for warpgroup acceleration capabilities via
smart uses of PTX-level instructions. }
Hopper Tensor Cores, increasingly introduce Warpgroup Matrix Multiply-Accumulate (\texttt{wgmma}) instruction. This instruction however imposes a key constraint: in a matrix multiplication $C = AB$, only $A$ and $C$ can be sourced from registers, while $B$ must reside in shared memory. This presents a challenge for low-bit quantized data, as values are typically upconverted to FP16 in registers before computation. To resolve this, we leverage Hopper's \texttt{STSM} PTX instruction to store dequantized FP16 values in shared memory efficiently, accessible for \texttt{wgmma\_SS} operations. Remarkably, the asynchronous nature of WGMMA overlaps storage with computation, optimizing performance.

\textbf{(2) Accelerating Blackwell with native low-precision format.}
The Blackwell architecture introduces native support for low-precision tensor operations, eliminating the need for explicit dequantization. Consequently, the \texttt{lop3}-based register remapping described earlier is bypassed in favor of direct execution. We target Blackwell’s low-precision \texttt{mma} instructions—specifically those supporting the micro-scaling formats (e.g., \texttt{mxfp4 /\ nvfp4})—to execute GEMM operations directly on packed 4-bit data. While these instructions enforce rigid layout constraints for both the packed values and their block-scaling factors, the layout transformation strategy proposed in Section~\ref{subsec:Scheme} is designed to be layout-agnostic. It automatically aligns the packed KV data with the hardware-mandated format, ensuring seamless integration with Blackwell’s native tensor pipelines.

\section{Evaluation}

In this section, we comprehensively evaluate BitDecoding against state-of-the-art approaches and systems. Our evaluation highlights the following key results:

\begin{enumerate}
\item BitDecoding outperforms FP16 FlashDecoding-v2 by significant margins across GPU generations, achieving speedups of up to 8.6$\times$ on Blackwell (using native MXFP4), 8.0$\times$ on Hopper, and 7.5$\times$ on Ada architectures, while surpassing the state-of-the-art low-bit system QServe by up to 4.3$\times$ (Section~\ref{sec:kernel-eval}).
\item In end-to-end long-context inference, BitDecoding reduces single-batch latency by 3x (on LLaMA-3.1-8B with 128K context) and achieves over 4x higher serving throughput than QServe, demonstrating superior scalability in GQA settings where prior CUDA Core-only methods degrade (Section~\ref{sec:end-to-end-eval}).
\item BitDecoding preserves near-FP16 accuracy while deriving significant performance gains from each system component, demonstrating only a 0.2\% accuracy degradation with 4-bit quantization, while our ablation study confirms that every design module contributes to the overall speedup (Section~\ref{sec:details-eval}).

\end{enumerate}




\subsection{Kernels Performance Across GPU Architectures}
\label{sec:kernel-eval}

\textbf{Kernels Settings.} Since different LLM serving scenarios require varying workloads and attention kernel designs, we evaluate performance under the following three representative settings:

\begin{itemize}  
    \item \textbf{Single:} A scenario where \( \text{batch\_size} = 1 \), representing inference for edge users with long context.  
    \item \textbf{Batches:} A setting with a larger \( \text{batch\_size} \), maintaining the same input length while applying simple padding.  
    \item \textbf{Page:} A high-throughput scenario where a larger \( \text{batch\_size} \) is managed using the page management technique~\cite{page}.
\end{itemize}


\textbf{Baselines.} We compare BitDecoding against several representative attention kernel implementations. For FP16 KV cache, we use FlashDecoding~\cite{dao2023flashattention2, fa3}—a split-partitioned variant of FlashAttention optimized for long-context decoding—as our baseline for speedup normalization. For low-bit KV cache, we evaluate Kivi~\cite{kivi}, a non-fused kernel supporting 4-bit and 2-bit quantization; Atom~\cite{atom} and QServe~\cite{qserve}, both fused-kernel implementations with CUDA Cores-only approach and supporting 4-bit cache with page management. Notably, Atom does not support GQA.

\textbf{Quantization Settings.} We evaluate BitDecoding under various quantization configurations, supporting 4-bit and 2-bit Key tensors with both Channel-wise (KC) and Tensor-wise (KT) schemes.

\begin{figure}[t]
    \centering
    \includegraphics[width=0.9\linewidth]{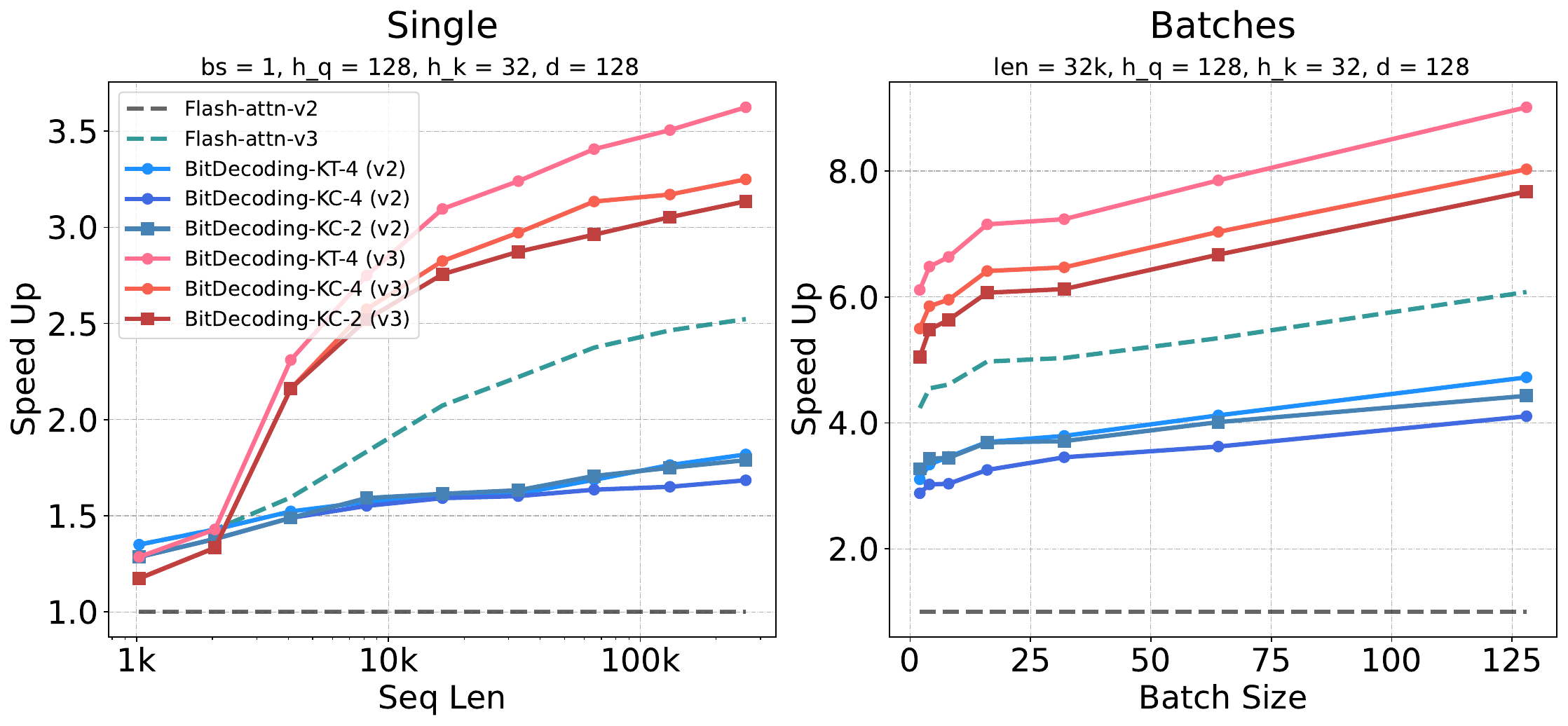}
    \caption{Kernel performance on Hopper (H100).}
    \label{fig:kernel_h100}
\end{figure}

\textbf{Results on MXFP4 /\ NVFP4 (RTX5090, RTX PRO 6000).} The Blackwell architecture provides native support for low-precision data formats, eliminating on-the-fly dequantization overhead while delivering very high compute throughput on low-bit operations. As shown in Fig.~\ref{fig:kernel_5090}, BitDecoding achieves remarkable performance, reaching up to 8.6$\times$ speedup in batched scenarios and over 4.3$\times$ in single-batch long-context decoding (128k), significantly outpacing the non-fused attention baseline. Similarly, Fig.~\ref{fig:kernel_6000} demonstrates that the RTX PRO 6000 attains substantial gains, peaking at 6.5$\times$ speedup with large batch sizes. 

\begin{figure*}[t]
    \centering
    \includegraphics[width=0.9\linewidth]{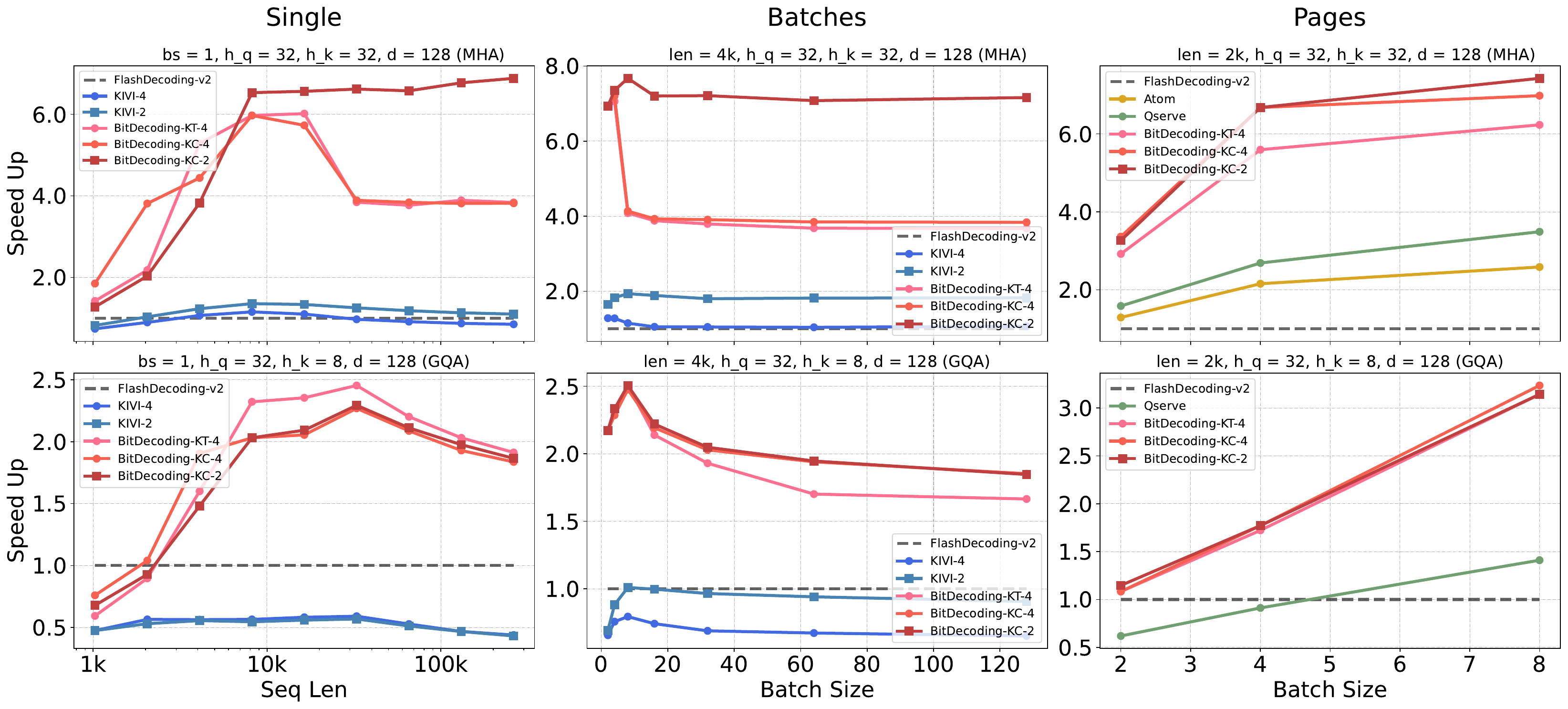}  
    \caption{Kernel performance on RTX4090.}
    \label{fig:kernel_4090}
\end{figure*}

\textbf{Results on Advanced Tensor Cores Acceleration (H100).} Newer GPU architectures often introduce advanced compute instructions that significantly accelerate kernel execution. As illustrated in Fig.~\ref{fig:kernel_h100}, FlashDecoding-v3, optimized for Hopper Tensor Cores, delivers notable performance gains over its v2 counterpart. While BitDecoding-v2 reaches up to 4.1$\times$ speedup, the v3 implementation further boosts performance to 8.0$\times$. This is enabled by BitDecoding's use of Hopper’s \texttt{wgmma} and asynchronous memory instructions, ensuring high Tensor Cores utilization even in mixed-precision settings.


\begin{figure*}[h]
    \centering
    \includegraphics[width=0.9\linewidth]{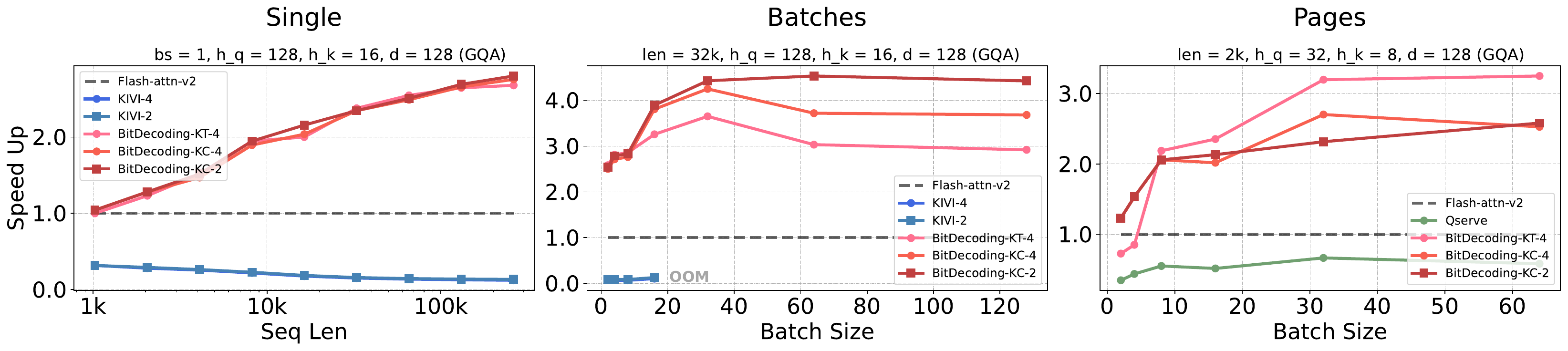}  
    \caption{Kernel performance on A100.}
    \label{fig:kernel_a100}
\end{figure*}

\textbf{Results on Bandwidth-constrained GPU (RTX 4090).} Leveraging low-precision data is critical for accelerating inference on bandwidth-constrained GPUs. As shown in Fig.~\ref{fig:kernel_4090}, BitDecoding achieves roughly $4\times$ (4-bit) and over $7\times$ (2-bit) speedups over FlashDecoding-v2 in Single and Batches settings, gains that stem directly from alleviating DRAM bottlenecks via low-bit KV caching. 

BitDecoding significantly outperforms baselines across all scenarios; unlike the non-fused KIVI, which relies on separate kernels and suffers severe degradation in GQA, BitDecoding’s fully fused design maintains high efficiency. In Page settings, it surpasses fused CUDA-core baselines: for MHA, BitDecoding achieves over $6\times$ speedup compared to QServe's $3.5\times$. Crucially, in compute-intensive GQA, it maintains a $3\times$ speedup while QServe drops to $1.4\times$, confirming that leveraging Tensor Cores provides robust acceleration where CUDA-only approaches falter.

\textbf{Results on High-Bandwidth GPU (A100).} On architectures with high memory bandwidth like the A100, computation pressure becomes more pronounced, as performance bottlenecks shift from memory access to compute utilization—especially when kernel designs fail to fully exploit available compute resources. As shown in Fig.~\ref{fig:kernel_a100}, both KIVI and QServe suffer from poor performance—KIVI due to its non-fused kernel design, and QServe due to underutilization of Tensor Cores—even performing worse than the FP16 baseline. In contrast, BitDecoding consistently outperforms all baselines across workloads, achieving up to $3\times$ speedup, thanks to its efficient utilization of Tensor Cores and fused execution pipeline. An interesting observation is that the performance gap between 4-bit and 2-bit variants narrows on A100, as the increased DRAM bandwidth reduces memory bottlenecks and shifts the performance balance toward compute-bound execution.

\subsection{Performance across LLMs Inference Systems}
\label{sec:end-to-end-eval}

\textbf{Model settings.} We evaluate on a range of LLMs, including LLaMA-2-7B, LLaMA-3.1-8B, LLaMA-3.1-70B, Qwen3-8B, and Qwen3-14B. Among them, only LLaMA-2-7B adopts MHA, while the others use GQA. All models are run on a single A100 GPU, except LLaMA-3.1-70B, which is evaluated on 8×A100 GPUs.

\textbf{Quantization settings.} We choose channel-wise quantization for LLMs KV cache as it brings better accuracy and aligns with the Kivi.

\begin{figure}[t]
    \centering
    \includegraphics[width=1\linewidth]{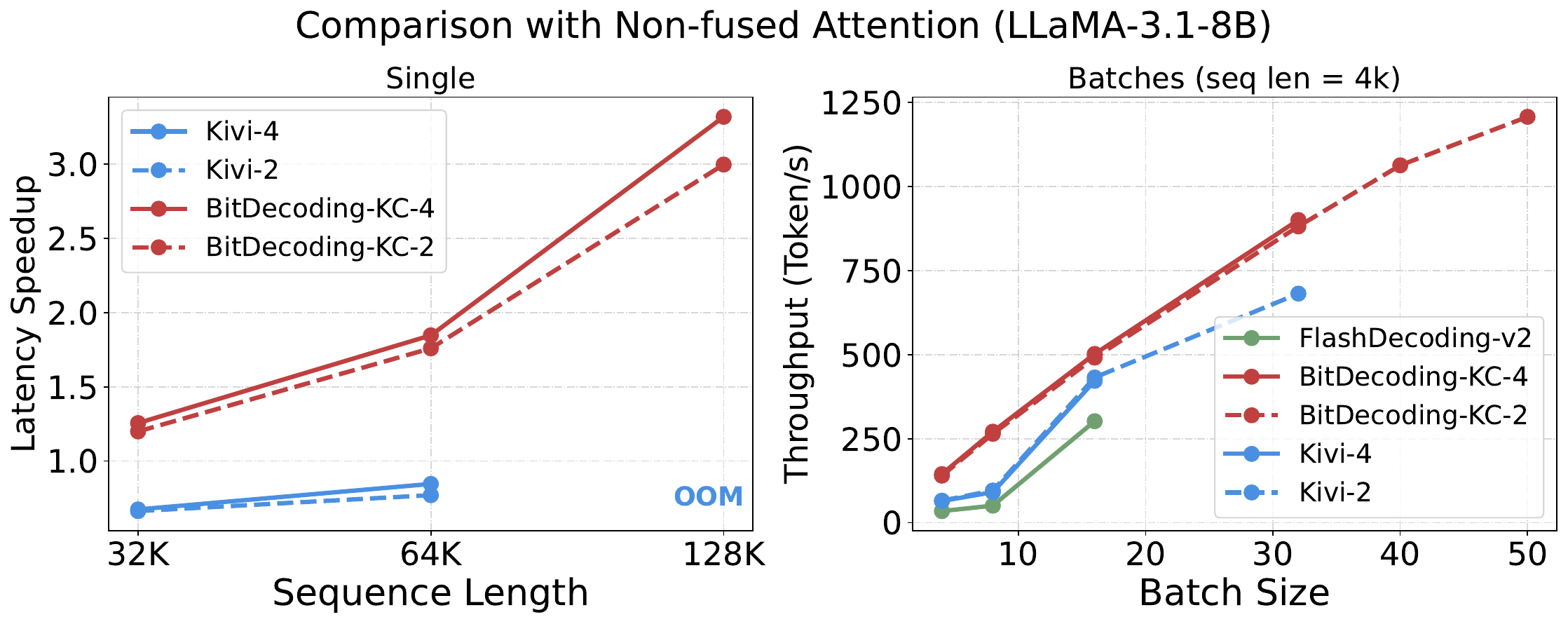}
    \caption{Comparing Kivi with (a) end-to-end generation time and (b) decoding throughput.}
    \label{fig:single}
\end{figure}

\textbf{Compared with Non-fused Attention.} As illustrated in Fig.~\ref{fig:single}, in the Single setting, BitDecoding achieves up to 3.3$\times$ speedup at a 128K context length, where KV cache loading becomes the dominant bottleneck in LLMs inference. In contrast, Kivi suffers from limited scalability and encounters out-of-memory (OOM) failures at 128K due to the lack of block-tiling kernel support. For the Batches setting, BitDecoding significantly outperforms KIVI in throughput: BitDecoding-KC-4 and KC-2 reach up to 900 and 1200 tokens/s, respectively, while KIVI-4 and KIVI-2 peak below 700 tokens/s.

\begin{figure}[t]
    \centering
    \includegraphics[width=0.97\linewidth]{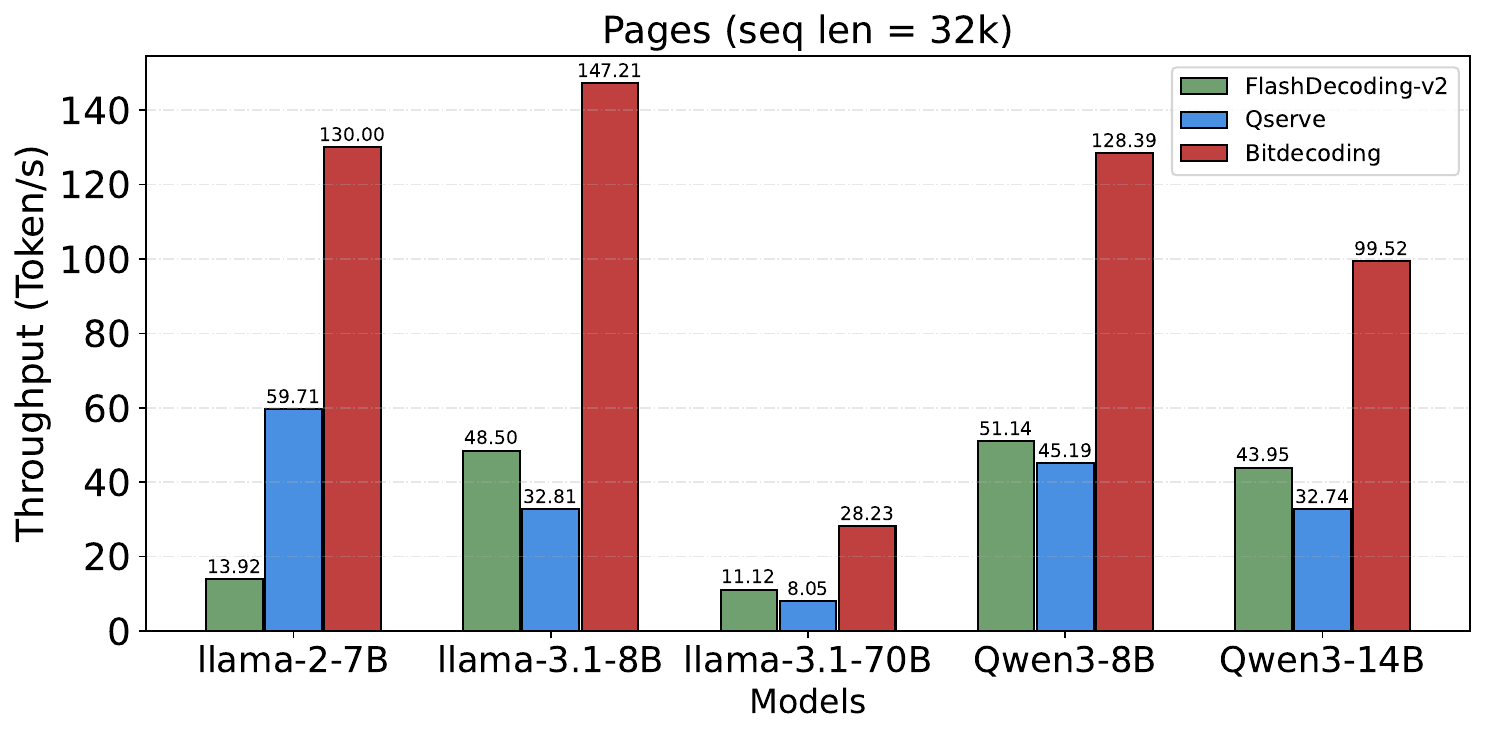}
    \caption{Comparing Qserve with decoding throughput.}
    \label{fig:page}
\end{figure}

\textbf{Compared with CUDA Cores-only fused Attention.} We compare BitDecoding with Qserve for page-setting inference, as Qserve supports both MHA and GQA attention structures. The maximum throughput is evaluated under the largest batch sizes available within GPU memory. As illustrated in Fig.~\ref{fig:page}, Qserve achieves higher throughput than FlashDecoding-v2 on LLaMA-2-7B but suffers from degraded performance on all other models due to inefficiencies in handling GQA. In contrast, BitDecoding consistently outperforms QServe across both LLaMA and Qwen architectures, under both single-GPU and multi-GPU settings, achieving more than 2$\times$ higher maximum throughput compared to QServe.

\subsection{Accuracy, Overhead and Performance Breakdown}
\label{sec:details-eval}

\textbf{Accuracy analysis.} As shown in Table~\ref{tab:bitwidth_comparison}, we evaluate throughput and accuracy across different bit widths. The 2-bit quantization reduces memory consumption significantly, enabling larger batch sizes and achieving a \( 4.25\times \) higher throughput compared to FP16. Meanwhile, the 4-bit quantization achieves a \( 2.98\times \) speedup while maintaining near full-precision accuracy with only a minimal \( 0.2\% \) degradation. These results highlight the trade-off, with 4-bit quantization offering balance and 2-bit maximizing throughput at a slight accuracy cost.

\begin{table}[htbp]
    \centering
    \caption{Efficiency and accuracy tradeoff with low-bit KV cache. We use Llama-3.1-8B-Instruct with $seq\_len=32K$, and evaluate average accuracy on longbench~\cite{bai2024longbench}.}
    
    \begin{tabularx}{\columnwidth}{XXX}
        \toprule
        \textbf{KV Cache} & \textbf{Throughput} & \textbf{Longbench Acc} \\
        \midrule
        FP16 & 49.25 & 48.25 \\
        INT4 & 147.21 (+2.98x) & 48.16 (-0.2\%) \\
        INT2 & 209.48 (+4.25x) & 47.38 (-2.7\%) \\
        \bottomrule
         \end{tabularx}
    \label{tab:bitwidth_comparison}
\end{table}

\begin{table}[h!]
\small
\centering
\caption{Latency (ms) comparison of quantization and packing during inference.}
\label{tab:quantization_inference}
\begin{tabular}{cccc}
\toprule
\textbf{Inference Phase} & \textbf{Marlin} & \textbf{Ladder} & \textbf{BitDecoding} \\
\midrule
Prefill & 58.02 & 4.79 & 0.0599 \\
Decode  & 0.41  & 0.65 & 0.008 \\
\bottomrule
\label{tab:quantization}
\end{tabular}
\end{table}

\begin{table}[h!]
    \centering
    \caption{Impact of cooperative softmax and warps on performance and validity.}
    \begin{tabular}{cccccc}
        \toprule
        \textbf{\textbf{$W_n$}} & \textbf{Coop. Soft} & \textbf{Latency (ms)} & \textbf{TCs Utilization (\%)} & \textbf{Valid} \\
        \midrule
        1 & \ding{53} & 3.746 & 10.91    & \checkmark \\
        4 & \ding{53} & 0.610 & 19.71 & \ding{53} \\
        4 & \checkmark & 0.613 & 19.66 & \checkmark \\
        \bottomrule
    \end{tabular}
    \label{tab:warp_softmax}
\end{table}

\begin{figure}[t]
    \centering
    \includegraphics[width=0.95\linewidth]{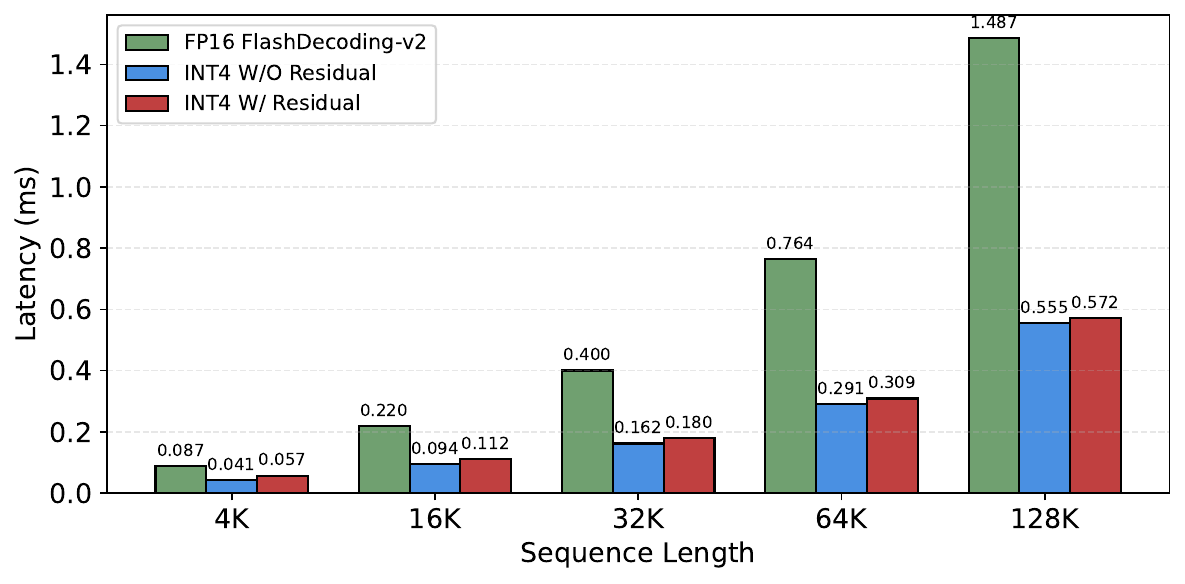}
    \caption{Runtime overhead of the residual KV cache.}
    \label{fig:residual}
\end{figure}

\textbf{Half-precision Residual Kernel Overhead.}
Half-precision residual KV Cache would introduce quite a small portion memory overhead as $seq\_len >> N_r$, while $seq\_len$ would be more than 32K and $N_r$ is always less than 256. The half-precision residual KV cache introduces only a slight runtime overhead due to an extra kernel launch, as shown in Fig.~\ref{fig:residual}. Moreover, this overhead becomes increasingly negligible as the sequence length grows, since the residual portion constitutes a smaller fraction of the total KV cache.

\textbf{Quantization and Packing Overhead.}  
We evaluate the latency of quantization and packing under a sequence length of $seq\_len=128K$, comparing BitDecoding with Marlin~\cite{marlin} and Ladder~\cite{ladder}. As shown in Table~\ref{tab:quantization}, the pre-transformation and packing step in previous mixed-precision computing methods introduce significant overhead, which cannot be ignored. Our kernel incurs minimal overhead after the Prefill phase, primarily due to kernel launch overhead. Moreover, during decoding, we achieves nearly negligible overhead, as it is fully fused into kernel computation.

\begin{figure}[t]
    \centering
    \begin{subfigure}[b]{0.49\linewidth}
        \centering
        \includegraphics[width=\linewidth]{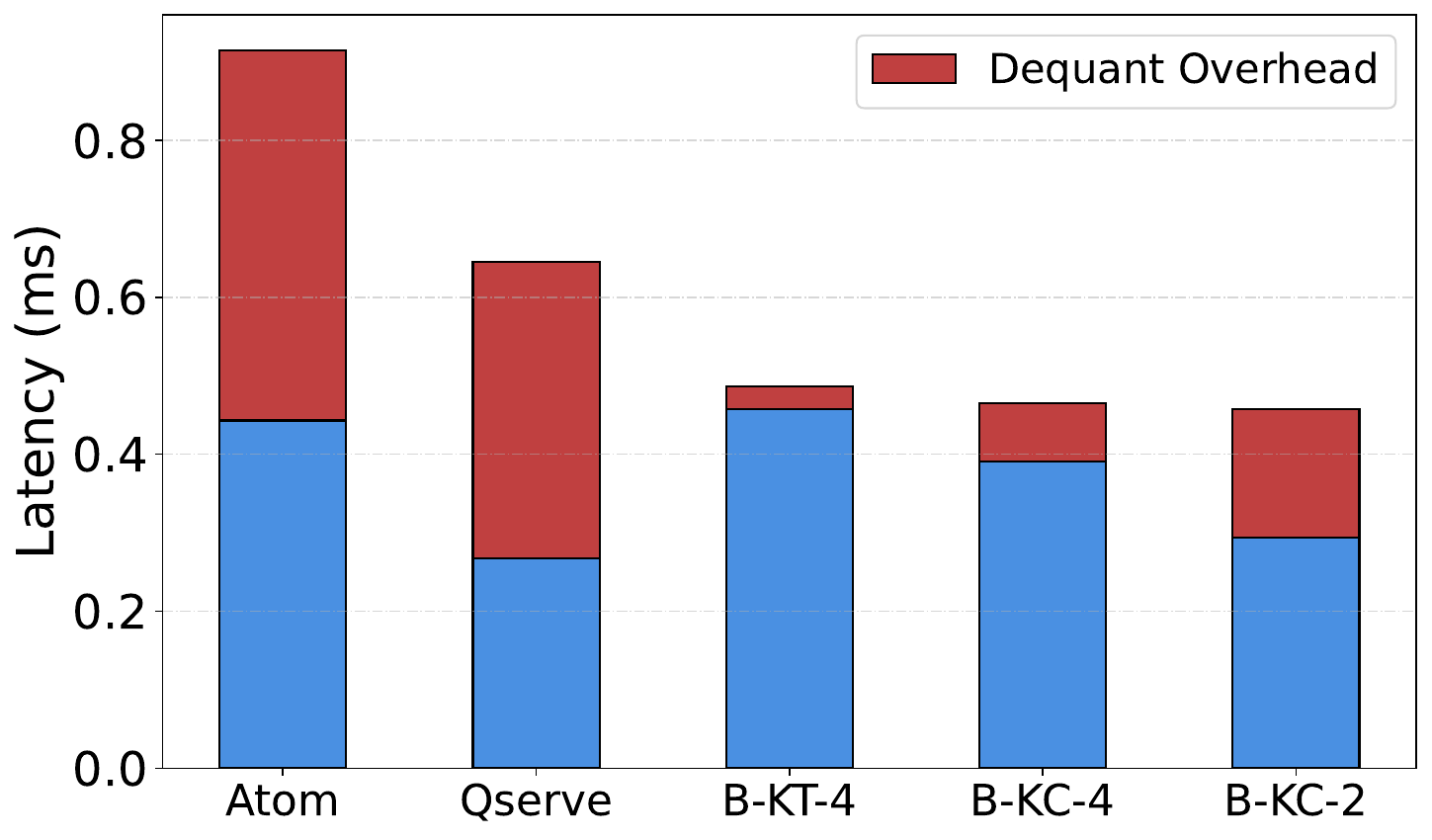}
        \caption{Dequantization Overhead}
        \label{fig:dequant}
    \end{subfigure}
    \hfill
    \begin{subfigure}[b]{0.49\linewidth}
        \centering
        \includegraphics[width=\linewidth]{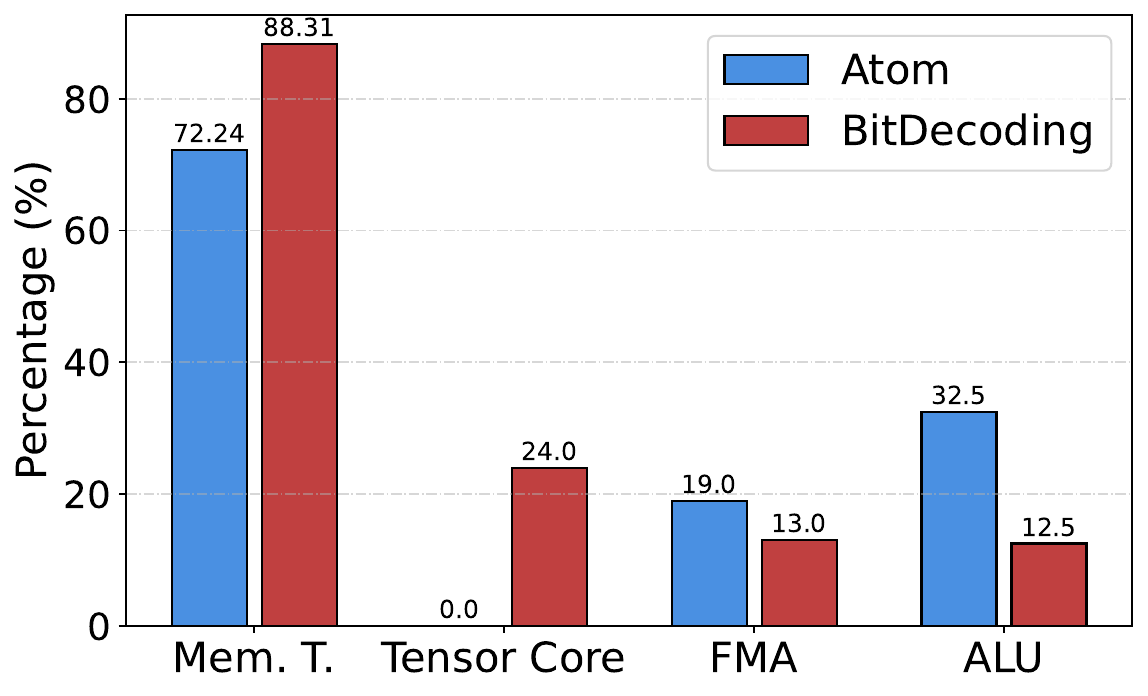}
        \caption{Micro Analysis}
        \label{fig:nsysc}
    \end{subfigure}
    \caption{Dequantization overhead analysis.}
    \label{fig:dequant_overhead}
\end{figure}

\textbf{Dequantization Overhead.} Fig.~\ref{fig:dequant} illustrates the high computational overhead of dequantization in Atom and QServe, consuming nearly half the kernel execution time. In contrast, BitDecoding significantly reduces this overhead to less than 15\% (4-bit) and 35\% (2-bit), thanks to better Tensor Cores overlap.

A further microbenchmark comparing Atom and BitDecoding (Fig.~\ref{fig:nsysc}) reveals BitDecoding's superior memory throughput from effective Tensor Core usage. Conversely, Atom relies heavily on CUDA cores, increasing pressure on FMA and ALU operations.

\textbf{Multi-warps Cooperative Softmax Overhead.} Table~\ref{tab:warp_softmax} shows that increasing $W_n$ improves Tensor Cores utilization and reduces latency, but breaks correctness without cooperative softmax. Enabling cooperative softmax restores correctness with only 0.5\% overhead. Although it introduces shared memory access, the overhead is minimal since low-bit data reduces memory bandwidth pressure and shifts the kernel from memory-bound to compute-bound.

\begin{figure}[t]
    \centering
    \includegraphics[width=0.9\linewidth]{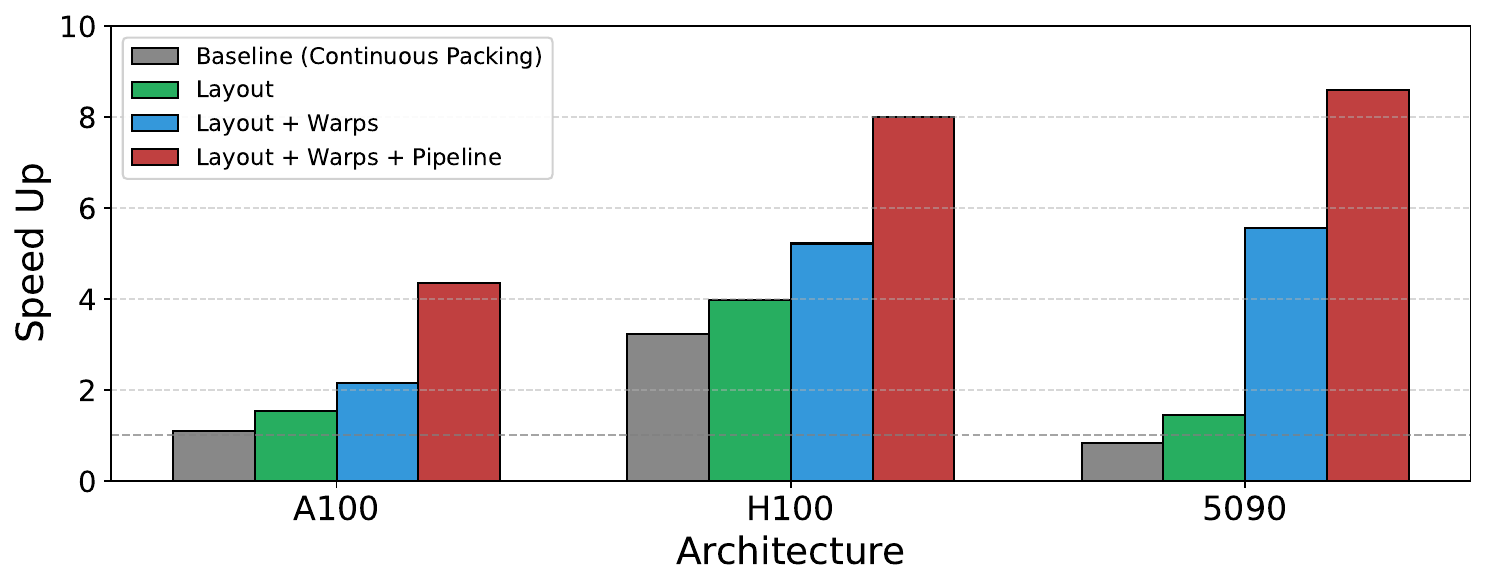}
    \caption{Breakdown of BitDecoding optimizations across architectural generations.}
    \label{fig:breakdown}
\end{figure}

\textbf{BreakDown Analysis.}
To further analyze the performance gains of BitDecoding, we decompose our optimizations in Fig.~\ref{fig:breakdown}. Following \cite{ashkboos2024quarot}, we use a continuous-packing baseline that quantizes and packs the KV cache at every generation step, which introduces substantial overhead and requires manual effort to maintain valid layouts. In contrast, our layout design automatically induces Tensor Core–compatible layouts for arbitrary low-bit formats, fully unlocking the compute potential of Tensor Cores. On top of this, the warp-parallelism strategy contributes significant additional speedups, while the pipeline optimizations further enhance end-to-end performance.

\section{Related Works}
\paragraph{KV Cache Quantization Algorithms} 
KV cache quantization reduces memory usage in LLMs with long contexts while maintaining performance.
Recent works explore 4-bit, 2-bit, and even 1-bit KV cache quantization, aiming to push the limits of compression.
Methods like KIVI~\cite{kivi}, Gear~\cite{gear}, and KVQuant~\cite{kvquant} use per-channel quantization to handle key-value outliers, while RotateKV~\cite{rotatekv} applies rotation to smooth channel-wise distributions. Although effective at higher compression ratios, these methods lack efficient system implementations, leading to suboptimal performance.

\paragraph{Mixed-precision Matrix Multiplication} Low-bit weight and low-bit KV cache in LLMs create a unique requirement for mixed-precision matrix multiplication (mpGEMM), where one input matrix is in lower precision (e.g., INT4/2/1) while the other matrix remains in higher precision (e.g., FP16/8). Optimized kernels like Ladder~\cite{ladder} and Marlin~\cite{marlin} improve performance via layout transformations and efficient dequantization. However, these methods require pre-packing and pre-transforming weights, limiting applicability to low-bit KV cache in autoregressive decoding.

\paragraph{System Implementation for Low-bit KV Cache} KIVI~\cite{triton} uses Triton with separate kernels for low-bit KV Cache implementation. Atom~\cite{atom} integrates quantization within the preceding linear layer, while QServe~\cite{qserve} fuses quantization directly into FlashAttention kernels. However, they both rely on GEMV operations with fused multiply–add (FMA) instructions, missing Tensor Core acceleration.

\section{Conclusion}

BitDecoding establishes a new system foundation for efficient low-bit KV-cache decoding by demonstrating how CUDA cores and Tensor Cores can be cooperatively orchestrated using principled system designs. Its layout-induction and warp-level coordination techniques generalize across attention variants, quantization schemes, and GPU generations, and naturally extend to emerging architectures such as Blackwell and even beyond. We expect BitDecoding to enable future work on algorithm–system co-design for KV-cache quantization, near-lossless test-time scaling, and more capable GPU execution models for long-context LLMs inference.

\bibliographystyle{IEEEtranS}
\bibliography{refs}

\end{document}